
\magnification \magstep1
\raggedbottom
\openup 4\jot
\voffset6truemm
\headline={\ifnum\pageno=1\hfill\else
\hfill{\it Mathematical Structures of Space-Time}
\hfill \fi}
\rightline {DAMTP R-90/5}
\vskip 1cm
\centerline {\bf MATHEMATICAL STRUCTURES OF SPACE-TIME$^{*}$}
\vskip 1cm
\centerline {\bf Giampiero Esposito}
\centerline {Department of Applied Mathematics and Theoretical Physics}
\centerline {Silver Street, Cambridge CB3 9EW, U. K.}
\centerline {and}
\centerline {St. John's College, Cambridge CB2 1TP, U. K.}
\vskip 1cm
\centerline {March-April 1990}
\vskip 1cm
\noindent
{\bf Abstract.}
At first we introduce the space-time manifold and we compare some aspects
of Riemannian and Lorentzian geometry such as the distance function and the
relations between
topology and curvature. We then define spinor structures
in general relativity, and the conditions for their existence are discussed.
The causality conditions are studied through an analysis
of strong causality, stable causality and global hyperbolicity. In looking
at the asymptotic structure of space-time, we focus on the
asymptotic symmetry group of Bondi, Metzner and Sachs, and the b-boundary
construction of Schmidt.
The Hamiltonian structure of space-time
is also analyzed, with emphasis on Ashtekar's spinorial variables.

Finally, the question of a rigorous theory of
singularities in space-times with torsion is addressed, describing in detail
recent work by the author. We define geodesics
as curves whose tangent vector moves by parallel transport. This is different
from what other authors do, because their
definition of geodesics only involves the
Christoffel symbols, though studying theories with torsion.
We then prove how to extend Hawking's singularity
theorem without causality assumptions to the space-time of
the ECSK theory. This is achieved studying the generalized Raychaudhuri
equation in the ECSK theory, the conditions for the existence of
conjugate points and properties of maximal timelike geodesics.
Our result can also be interpreted as a no-singularity theorem if the
torsion tensor does not obey some additional conditions.
Namely, it seems that the occurrence of singularities in closed
cosmological models based on the ECSK theory is less generic than in
general relativity. Our work should be compared with important previous
papers. There
are some relevant differences, because we rely on a different definition of
geodesics, we keep the field equations of the ECSK theory in their original
form rather than casting them in a form similar to general relativity with
a modified energy-momentum tensor, and we emphasize the role played by the
full extrinsic curvature tensor and by the variation formulae.
\vskip 5cm
\noindent
$^{*}$Fortschritte der Physik, {\bf 40}, 1-30 (1992).
\vskip 40cm
\leftline {\bf 1. INTRODUCTION}
\vskip 1cm
The space-time manifold plays still a vital role in modern relativity
theory, and we are going to examine it in detail through an analysis
of its mathematical structures. Our first aim is to present an unified
description of some aspects of Lorentzian and Riemannian geometry, of the
theory of spinors, and of causal, asymptotic and Hamiltonian structure.
This review paper is aimed both at theoretical and mathematical physicists
interested in relativity and gravitation,
and it tries to present together several topics which
are treated in greatly many more books and original papers. Thus in
section 2, after defining the space-time manifold, following [1]
we discuss the distance function and the relations between
topology and curvature in Lorentzian and Riemannian geometry. In
section 3 we use two-component spinor language which is more familiar to
relativists. At first we define spin space and Infeld-Van Der Waerden
symbols, and then we present the results of Geroch on spinor structures.
This section can be seen in part as complementary to important recent work
appeared in [2], and we hope it can help in improving the understanding of
the foundational points of a classical treatise such as [3].
In section 4, after some basic definitions, we study
three fundamental causality conditions such as strong causality, stable
causality and global hyperbolicity. In section 5, the asymptotic structure
of space-time is studied focusing on the asymptotic symmetry group of
Bondi-Metzner-Sachs (hereafter referred to as BMS) and on the boundary
of space-time. This choice of arguments is motivated by the second part of
our paper, where the singularity theory in cosmology for space-times with
torsion is studied. In fact the Poincar\'e group can be seen as the
subgroup of the BMS group which maps good cuts into good cuts, and it is
also known that the gauge theory of the Poincar\'e group leads to theories
with torsion [4-7]. Thus it appears important to
clarify these properties. The boundary of space-time is studied in section
5.2 defining the b-boundary of Schmidt [8], discussing its construction
and related questions [9].
In section 6 we present Ashtekar's spinorial
variables for canonical gravity [10]. In agreement with the aims of
our paper, we only emphasize the classical aspects of Ashtekar's theory.
This section presents a striking application of the concepts defined in
section 3, and it illustrates the modern approach to the Hamiltonian
formulation of general relativity. Finally, section 7 is devoted to the
clarification of recent work of the author [11] on the
singularity problem for space-times with torsion, using also concepts
defined in sections 2, 4 and 5.

So far, the singularity problem for theories with torsion had been studied
defining geodesics as extremal curves. However, a rigorous theory of
geodesics in general relativity can be based on the concept of autoparallel
curves [1,12]. Thus it appears
rather important to develop the mathematical theory of singularities
when geodesics are defined as curves whose tangent vector moves by
parallel transport. This definition involves the full connection with
torsion, whereas extremal curves just involve the Christoffel symbols.
In so doing one appreciates the role of the full extrinsic curvature
tensor and of the variation formulae, two important concepts which were not
considered in [13]. One can also see that one can
keep the field equations of the Einstein-Cartan-Sciama-Kibble
(hereafter referred to as ECSK) theory in their original form, rather
than casting them (as done in [13]) in a form similar to
general relativity but with a modified energy momentum tensor. We then
follow and clarify [11] in proving how to extend Hawking's
singularity theorem without causality assumptions to the space-time of
the ECSK theory. In the end, our concluding remarks are presented in
section 8.
\vskip 1cm
\leftline {\bf 2. LORENTZIAN AND RIEMANNIAN GEOMETRY}
\vskip 1cm
\leftline {\bf 2.1. The space-time manifold}
\vskip 1cm
A space-time $(M,g)$ is the following collection of mathematical entities
[1,12] :

(1) A connected four-dimensional Hausdorff $C^{\infty}$ manifold $M$ ;

(2) A Lorentz metric $g$ on $M$, namely the assignment of a nondegenerate
bilinear form $g_{\mid p}:T_{p}M$x$T_{p}M\rightarrow R$ with diagonal form
$(-,+,+,+)$ to each tangent space. Thus $g$ has signature $+2$ and is not
positive-definite ;

(3) A time orientation, given by a globally defined timelike vector field
$X : M\rightarrow TM$. A timelike or null tangent vector $v\in T_{p}M$ is
said to be future-directed if $g(X(p),v)<0$, or past-directed if
$g(X(p),v)>0$.

Some important remarks are now in order :

(a) The condition (1) can be formulated for each number of space-time
dimensions $\geq 2$ ;

(b) Also the convention $(+,-,-,-)$ for the diagonal form of the metric
can be chosen [14]. This convention seems to be more useful in
the study of spinors, and can be adopted also in using tensors as Penrose
does so as to avoid a change of conventions. The definitions of timelike
and spacelike will then become opposite to our definitions :
$X$ is timelike if $g(X(p),X(p))>0$ $\forall p \in M$, and $X$ is spacelike
if $g(X(p),X(p))<0$ $\forall p \in M$ ;

(c) The pair $(M,g)$ is only defined up to equivalence. Two pairs $(M,g)$
and $(M',g')$ are equivalent if there is a diffeomorphism
$\alpha : M\rightarrow M'$ such that : $\alpha_{*}g=g'$. Thus we are
really dealing with an equivalence class of pairs [12].

The fact that the metric is not positive-definite is the source of several
mathematical problems. This is why mathematicians generally focused their
attention on Riemannian geometry. We are now going to sum up some basic
results of Riemannian geometry, and to formulate their counterpart
(when possible) in Lorentzian geometry. This comparison is also very useful
for gravitational physics. In fact Riemannian geometry is related to the
Euclidean path-integral approach to quantum gravity [15],
whereas Lorentzian geometry is the framework of general relativity.
\vskip 1cm
\leftline {\bf 2.2. Riemannian geometry versus Lorentzian geometry}
\vskip 1cm
A Riemannian metric $g_{0}$ on a manifold $M$ is a smooth and
positive-definite section of the bundle of symmetric bilinear 2-forms on $M$.
A fundamental result in Riemannian geometry is the Hopf-Rinow theorem. It
can be formulated as follows [1] :
\vskip 0.3cm
\noindent
{\bf Theorem 2.1.} : For any Riemannian manifold $(M,g_{0})$ the following
properties are equivalent :

(1) Metric completeness : $M$ together with the Riemannian distance function
(see section 2.2.1.) is a complete metric space ;

(2) Geodesic completeness : $\forall v \in TM$, the geodesic $c(t)$ in $M$
such that $c'(0)=v$ is defined $\forall t \in R$ ;

(3) For some $p\in M$, the exponential map $exp_{p}$ is defined on the
entire tangent space $T_{p}M$ ;

(4) Finite compactness : any subset $K$ of $M$ such that
$sup \left \{d_{0}(p,q) : p,q \; \in K \right \} <\infty$ has compact
closure.

Moreover, if any of these properties holds, we also know that :

(5) $\forall p,q \in M$, there exists a smooth geodesic segment $c$ from
$p$ to $q$ with $L_{0}(c)=d_{0}(p,q)$ (namely any two points can be joined
by a minimal geodesic).

In Lorentzian geometry there is no sufficiently strong analogue to the
Hopf-Rinow theorem. However, one can learn a lot comparing the definitions
of distance function and the relations between topology and curvature in
the two cases.
\vskip 1cm
\leftline {\bf 2.2.1. The distance function in Riemannian geometry}
\vskip 1cm
Let $\Omega_{pq}$ be the set of piecewise smooth curves in $M$ from $p$
to $q$. Given $c : [0,1]\rightarrow M$ and belonging to $\Omega_{pq}$,
there is a finite partition of $[0,1]$ such that $c$ restricted to the
sub-interval $[t_{i},t_{i+1}]$ is smooth $\forall i$. The Riemannian arc
length of $c$ with respect to $g_{0}$ is defined by :
$$
L_{0}(c)\equiv \sum_{i=1}^{k-1}\int_{t_{i}}^{t_{i+1}}
\sqrt{g_{0}(c'(t),c'(t))} \; dt \; \; \; \; .
\eqno (2.1)
$$
The Riemannian distance function $d_{0} : M$x$M\rightarrow [0,\infty)$ is then
defined by [1] :
$$
d_{0}(p,q)\equiv \inf \left \{L_{0}(c):c\in \Omega_{pq} \right \}
\; \; \; \; .
\eqno (2.2)
$$
Thus $d_{0}$ has the following properties :

(1) $d_{0}(p,q)=d_{0}(q,p) \; \; \; \; \forall p,q \in M$ ;

(2) $d_{0}(p,q)\leq d_{0}(p,r)+d_{0}(r,q) \; \; \; \; \forall p,q,r \in M$ ;

(3) $d_{0}(p,q)=0$ if and only if $p=q$ ;

(4) $d_{0}$ is continuous and, $\forall p \in M$ and $\epsilon >0$, the
family of metric balls $B(p,\epsilon)=\left\{q\in M : d_{0}(p,q)<\epsilon
\right \}$ is a basis for the manifold topology.
\vskip 1cm
\leftline {\bf 2.2.2. The distance function in Lorentzian geometry}
\vskip 1cm
Let $\Omega_{pq}$ be the space of all future-directed nonspacelike curves
$\gamma : [0,1]\rightarrow M$ with $\gamma(0)=p$ and $\gamma(1)=q$. Given
$\gamma \in \Omega_{pq}$ we choose a partition of $[0,1]$ such that $\gamma$
restricted to $[t_{i},t_{i+1}]$ is smooth $\forall i=0,1,...,n-1$. The
Lorentzian arc length is then defined as [1] :
$$
L(\gamma)\equiv \sum_{i=0}^{n-1}\int_{t_{i}}^{t_{i+1}}
\sqrt{-g(\gamma'(t),\gamma'(t))} \; dt \; \; \; \; .
\eqno (2.3)
$$
The Lorentzian distance function $d : M$x$M\rightarrow R \cup\{\infty\}$
is thus defined as follows. Given $p\in M$, if $q$ does not belong to the
causal future of $p$ (see section 4) : $q \notin J^{+}(p)$, we set
$d(p,q)=0$. Otherwise, if $q\in J^{+}(p)$, we set [1] :
$$
d(p,q)\equiv sup \left \{L_{g}(\gamma) : \gamma \in \Omega_{pq} \right \}
\; \; \; \; .
\eqno (2.4)
$$
Thus such $d(p,q)$ may not be finite, if timelike curves from $p$ to $q$
attain arbitrarily large arc lengths. It also fails to be symmetric in
general, and one has : $d(p,q) \geq d(p,r)+d(r,q)$ if there are
future-directed
nonspacelike curves from $p$ to $r$ and from $r$ to $q$. Finally,
we need to recall the definition of timelike diameter $diam (M,g)$ of a
space-time $(M,g)$ [1] :
$$
diam(M,g)\equiv sup \left \{d(p,q) : p,q \in M \right \}
\; \; \; \; .
\eqno (2.5)
$$
\vskip 1cm
\leftline {\bf 2.2.3. Topology and curvature in Riemannian geometry}
\vskip 1cm
A classical result is the Myers-Bonnet theorem which
shows how the properties of the Ricci curvature may influence the topological
properties of the manifold. In fact one has [16] :
\vskip 0.3cm
\noindent
{\bf Theorem 2.2.} : Let $(M,g)$ be a complete $n$-dimensional Riemannian
manifold with Ricci curvature $Ric(v,v)$ such that :
$Ric(v,v)\geq {(n-1)\over r}$. Then $diam(M,g)\leq diam (S^{n}(r))$,
$diam(M,g)\leq \pi \sqrt{r}$, and $M$ is compact. Moreover, $M$ has finite
fundamental homotopy group.
\vskip 1cm
\leftline {\bf 2.2.4. Topology and curvature in Lorentzian geometry}
\vskip 1cm
The Lorentzian analogue of the Myers-Bonnet theorem can be formulated in the
following way [1] :
\vskip 0.3cm
\noindent
Let $(M,g)$ be a $n$-dimensional globally hyperbolic space-time (see
section 4.3.) such that either :

(1) All timelike sectional curvatures are $\leq -l <0$, or :

(2) $Ric(v,v)\geq (n-1)l>0$ $\forall$ unit timelike vectors $v\in TM$.
\vskip 0.3cm
\noindent
Then $diam(M,g)\leq {\pi \over \sqrt{l}}$.

The proof of this theorem, together with the discussion of the Lorentzian
analogue of the index and Rauch I,II comparison theorems can be found in
chapter $10$ of [1]. For another recent treatise on Riemannian geometry,
see [17]. An enlightening comparison of Riemannian and Lorentzian
geometry can also be found in [18].
\vskip 1cm
\leftline {\bf 3. SPINOR STRUCTURE}
\vskip 1cm
A full account of two-component spinor calculus may be found in
[3,19-20]. Here we just wish to recall the following definitions.

Spin space [21]
is a pair $(\Sigma,\epsilon)$, where $\Sigma$ is a two-dimensional
vector space over the complex or real numbers and $\epsilon$ a symplectic
structure on $\Sigma$. Such an $\epsilon$ provides an isomorphism between
$\Sigma$ and the dual space $\Sigma^{*}$. One has : $\lambda^{A}\in \Sigma$,
$\lambda_{A}\in \Sigma^{*}$. Unprimed (primed) spinor indices take the
values $0$ and $1$ ($0'$ and $1'$). They can be raised and lowered by
means of $\epsilon^{AB}$, $\epsilon_{AB}$, $\epsilon^{A'B'}$,
$\epsilon_{A'B'}$, which are all given by :
$\pmatrix{ 0&1 \cr -1&0 \cr}$, according to the rules :
$\rho^{A}=\epsilon^{AB}\rho_{B}$, $\rho_{A}=\rho^{B}\epsilon_{BA}$,
$\rho^{A'}=\epsilon^{A'B'}\rho_{B'}$, $\rho_{A'}=\rho^{B'}
\epsilon_{B'A'}$. An isomorphism exists between the tangent space $T$ at
a point of space-time and the tensor product of the unprimed spin space
$S$ and the primed spin space $S'$ : $T \cong S \otimes S'$. The
Infeld-Van Der Waerden symbols $\sigma_{\; \; AA'}^{a}$ and
$\sigma_{a}^{\; \; AA'}$ express this isomorphism, and the correspondence
between a vector $v^{a}$ and a spinor $v^{AA'}$ is given by [22] :
$$
v^{AA'}=\sigma_{a}^{\; \; AA'}v^{a} \; \; \; \; ,
\eqno (3.1)
$$
$$
v^{a}=\sigma_{\; \; AA'}^{a}v^{AA'} \; \; \; \; .
\eqno (3.2)
$$
The $\sigma_{a}^{\; \; AA'}$ are given by :
$$
\sigma_{0}=-{I\over \sqrt{2}} \; \; \; \; , \; \; \; \;
\sigma_{i}={\Sigma_{i}\over \sqrt{2}} \; \; \; \; ,
\eqno (3.3)
$$
where $\Sigma_{i}$ are the Pauli matrices. We are now going to focus
our attention on some more general aspects, following [23].

In defining spinors at a point of space-time, we may start by addressing
the question of how an array of complex numbers $\mu_{CD'}^{AB'}$ gets
transformed in going from a tetrad $v$ at $p$ to a tetrad $w$ at $p$.
The mapping $L :v\rightarrow w$ between $v$ and $w$ is realized by an
element $L$ of the restricted Lorentz group $L_{0}$ (so that it preserves
temporal direction and spatial parity). Now, to each $L$ there correspond
two elements $\pm U_{\; \; B}^{A}$ of $SL(2,C)$. Thus the transformation
law contains a sign ambiguity :
$$
\mu_{CD'}^{AB'}(w)=\pm U_{\; \; E}^{A}{\overline U}_{\; \; F'}^{B'}
{\left(U^{-1}\right)}_{\; \; C}^{G}
{\overline {U^{-1}}}_{\; \; D'}^{H'}
\mu_{GH'}^{EF'}(v) \; \; \; \; .
$$
So as to remove this sign ambiguity, let us consider the six-dimensional
space : $\psi \equiv \Bigr \{$ set of all tetrads at $p$ $\bigr \}$. We then
move to the universal covering manifold ${\widetilde \psi}$ of $\psi$ :
$$
{\widetilde \psi}\equiv \left \{(v,\alpha) : v \in \psi, \alpha=path \; in
\; \psi \; from \; v \; to \; w \right \} \; \; \; \; .
$$
{\bf Definition 3.1.} $(v,\alpha)$ is equivalent to $(u,\beta)$ if $u=v$ and
if we can continuously deform $\alpha$ into $\beta$ keeping fixed the terminal
points.

An important property (usually described by the Dirac scissors argument) is
that the tetrad at $p$ changes after a $2\pi$ rotation, but gets unchanged
after a $4\pi$ rotation. The advantage of considering ${\widetilde \psi}$
is that in so doing, $\forall v,w \in {\widetilde \psi}$, there is an unique
element $U_{\; \; B}^{A}$ of $SL(2,C)$ which transforms $v$ into $w$. Thus
we give [23] :
\vskip 0.3cm
\noindent
{\bf Definition 3.2.} A spinor at $p$ is a rule which assigns to each
$v \in {\widetilde \psi}$ an array $\mu_{CD'}^{AB'}$ of complex numbers such
that, given $v,w \in {\widetilde \psi}$ related by $U_{\; \; B}^{A} \in
SL(2,C)$, then :
$$
\mu_{CD'}^{AB'}(w)=U_{\; \; E}^{A}{\overline U}_{\; \; F'}^{B'}
{\left(U^{-1}\right)}_{\; \; C}^{G}
{\overline {U^{-1}}}_{\; \; D'}^{H'}
\mu_{GH'}^{EF'}(v) \; \; \; \; .
\eqno (3.4)
$$
In defining spinor structures on $M$, we start by considering :
$B=$ principal fibre bundle of oriented orthonormal tetrads on $M$. The
structure group of $B$ is the restricted Lorentz group, and the fibre at
$p\in M$ is the collection $\psi$ of tetrads at $p$ with given temporal
and spatial orientation. The sign ambiguity is corrected taking a fibre
bundle whose fibre is the universal covering space ${\widetilde \psi}$ [23].
\vskip 0.3cm
\noindent
{\bf Definition 3.3.} A spinor structure on $M$ is a principal fibre bundle
${\widetilde B}$ on $M$ with group $SL(2,C)$, together with a $2-1$
application $\phi : {\widetilde B}\rightarrow B$ such that :

(1) $\phi$ realizes the mapping of each fibre of ${\widetilde B}$ into a
single fibre of $B$ ;

(2) $\phi$ commutes with the group operations. Namely, $\forall U \in
SL(2,C)$ we have : $\phi U =E(U)\phi$ where $E:SL(2,C)\rightarrow L_{0}$
is the covering group of $L_{0}$.
\vskip 0.3cm
\noindent
{\bf Definition 3.4.} A spinor field on $M$ is a mapping $\mu$ of
${\widetilde B}$ into arrays of complex numbers such that (3.4) holds.

The basic theorems about spinor structures are the following [23] :
\vskip 0.3cm
\noindent
{\bf Theorem 3.1.} If a space-time $(M,g)$ has a spinor structure, for this
structure to be unique $M$ must be simply connected.
\vskip 0.3cm
\noindent
{\bf Theorem 3.2.} A space-time $(M,g)$ oriented in space and time has spinor
structure if and only if the second Stiefel-Whitney class vanishes.

{\bf Remark} : Stiefel-Whitney classes $w_{i}$ can be defined
for each vector bundle $\xi$ by means of a sequence of cohomology
classes $w_{i}(\xi)\in H^{i}(B(\xi);Z_{2})$. In so doing, we denote by
$H^{i}(B(\xi);Z_{2})$ the i-th singular cohomology group of $B(\xi)$ with
coefficients in $Z_{2}$, the group of integers modulo $2$ [24].
If $w_{2}\not =0$, one cannot define parallel
transport of spinors on $M$. The orientability of space-time assumed in
theorem 3.2. and in section 2.1. implies that also the first Stiefel-Whitney
class must vanish.
\vskip 0.3cm
\noindent
{\bf Theorem 3.3.} $(M,g)$ has a spinor structure if and only if the
fundamental homotopy groups of $B$ and $M$ are related by :
$$
\pi_{1}(B)\approx \pi_{1}(M) \oplus \pi_{1}(\psi)
=\pi_{1}(M)\oplus Z_{2} \; \; \; \; .
\eqno (3.5)
$$
\vskip 0.3cm
\noindent
{\bf Theorem 3.4.} A space-time $(M,g)$ space and time-oriented has spinor
structure if and only if each of its covering manifolds has spinor structure.
\vskip 0.3cm
\noindent
{\bf Theorem 3.5.} Let $M$ be noncompact. Then $(M,g)$ has spinor structure
if and only if a global system of orthonormal tetrads exists on $M$ [23].

When we unwrap $\psi$, we annihilate $\pi_{1}(\psi)$. The existence of a
spinor structure implies we can unwrap all fibres on $B$. Spinor structures
are related to the second homotopy group of $M$, whereas covering spaces are
related to the first homotopy group. However, it is wrong to think that a
spinor structure can be created simply by taking a covering manifold.
In a space-time $(M,g)$ which does not have spinor structure, there must be
some closed curve $\gamma$ which lies in the fibre over $p \in M$
such that [23] :

(a) $\gamma$ is not homotopically zero in the fibre ;

(b) $\gamma$ can be contracted to a point in the whole bundle of frames.

A very important application of the spinorial formalism in general relativity
will be studied in section 6, where we define Ashtekar's spinorial
variables for canonical gravity.
\vskip 1cm
\leftline {\bf 4. CAUSAL STRUCTURE}
\vskip 1cm
Let $(M,g)$ be a space-time, and let $p\in M$. The chronological future
of $p$ is defined as [1,12] :
$$
I^{+}(p)\equiv \left \{q\in M : p<<q \right \}
\; \; \; \; ,
\eqno (4.1)
$$
namely $I^{+}(p)$ is the set of all points $q$ of $M$ such that there is a
future-directed timelike curve from $p$ to $q$. Similarly, we define the
chronological past of $p$ :
$$
I^{-}(p)\equiv \left \{q\in M : q<<p \right \} \; \; \; \; .
\eqno (4.2)
$$
The causal future of $p$ is then defined by :
$$
J^{+}(p)\equiv \left \{q\in M : p\leq q \right \} \; \; \; \; ,
\eqno (4.3)
$$
and similarly for the causal past :
$$
J^{-}(p)\equiv \left \{q\in M : q\leq p \right \} \; \; \; \; ,
\eqno (4.4)
$$
where $a\leq b$ means there is a future-directed nonspacelike curve from
$a$ to $b$. The causal structure of $(M,g)$ is the collection of past and
future sets at all points of $M$ together with their properties.
Following [19] and [25], we shall here recall the following
definitions, which will then be useful in section 4.3. and for further
readings.
\vskip 0.3cm
\noindent
{\bf Definition 4.1.} A set $\Sigma$ is achronal if no two points of
$\Sigma$ can be joined by a timelike curve.
\vskip 0.3cm
\noindent
{\bf Definition 4.2.} A point $p$ is an endpoint of the curve $\lambda$ if
$\lambda$ enters and remains in any neighbourhood of $p$.
\vskip 0.3cm
\noindent
{\bf Definition 4.3.} Let $\Sigma$ be a spacelike or null achronal
three-surface in $M$. The future Cauchy development (or future domain of
dependence) $D^{+}(\Sigma)$ of $\Sigma$ is the set of points $p \in M$
such that every past-directed timelike curve from $p$ without past
endpoint intersects $\Sigma$.
\vskip 0.3cm
\noindent
{\bf Definition 4.4.} The past Cauchy development $D^{-}(\Sigma)$ of
$\Sigma$ is defined interchanging future and past in definition 4.3. The
total Cauchy development of $\Sigma$ is then given by
$D(\Sigma)=D^{+}(\Sigma)\cup D^{-}(\Sigma)$.
\vskip 0.3cm
\noindent
{\bf Definition 4.5.} The future Cauchy horizon $H^{+}(\Sigma)$ of $\Sigma$
is given by :
$$
H^{+}(\Sigma)\equiv \left \{ X: X \in D^{+}(\Sigma), I^{+}(X)\cap
D^{+}(\Sigma)= \phi \right \} \; \; \; \; .
\eqno (4.5)
$$
Similarly, the past Cauchy horizon $H^{-}(\Sigma)$ is defined as :
$$
H^{-}(\Sigma)\equiv \left \{X : X \in D^{-}(\Sigma), I^{-}(X)\cap D^{-}
(\Sigma) = \phi \right \} \; \; \; \; .
\eqno (4.6)
$$
\vskip 0.3cm
\noindent
{\bf Definition 4.6.} The edge of an achronal set $\Sigma$ is given by all
points $p \in {\overline \Sigma}$ such that any neighbourhood $U$ of $p$
contains a timelike curve from $I^{-}(p,U)$ to $I^{+}(p,U)$ that does not
meet $\Sigma$ [18].

Our definitions of Cauchy developments differ indeed from the ones in
[12], in that Hawking and Ellis look at past-inextendible
curves which are timelike or null, whereas we agree with Penrose and
Geroch in not including null curves in the definition. We are now
going to discuss three fundamental causality conditions : strong causality,
stable causality and global hyperbolicity.
\vskip 1cm
\leftline {\bf 4.1. Strong causality}
\vskip 1cm
The underlying idea for the definition of strong causality is that there
should be no point $p$ such that every small neighbourhood of $p$ intersects
some timelike curve more than once [26]. Namely, the space-time
$(M,g)$ does not "almost contain" closed timelike curves. In rigorous
terms, strong causality is defined as follows [19] :
\vskip 0.3cm
\noindent
{\bf Definition 4.7.} Strong causality holds at $p \in M$ if
arbitrarily small neighbourhoods of $p$ exist which
each intersect no timelike curve in a disconnected set.

A very important characterization of strong causality can be given by
defining at first the Alexandrov topology [12].
\vskip 0.3cm
\noindent
{\bf Definition 4.8.} In the Alexandrov topology, a set is open if and only if
it is the union of one or more sets of the form :
$I^{+}(p)\cap I^{-}(q)$, $p,q \in M$.

Thus any open set in the Alexandrov topology will be open in the manifold
topology. Now, the following fundamental result holds [14] :
\vskip 0.3cm
\noindent
{\bf Theorem 4.1.} The following three requirements on a space-time $(M,g)$
are equivalent :

(1) $(M,g)$ is strongly causal ;

(2) the Alexandrov topology agrees with the manifold topology ;

(3) the Alexandrov topology is Hausdorff.
\vskip 1cm
\leftline {\bf 4.2. Stable causality}
\vskip 1cm
Strong causality is not enough to ensure that space-time is not just about
to violate causality [26]. The situation can be considerably
improved if stable causality holds. For us to be able to properly define
this concept, we must discuss the problem of putting a topology on the space
of all Lorentz metrics on a four-manifold $M$. Essentially three possible
topologies seem to be of major interest [26] : compact-open
topology, open topology, fine topology.
\vskip 0.3cm
\leftline {\bf 4.2.1. Compact-Open topology}
\vskip 0.3cm
$\forall i=0,1,...,r$, let $\epsilon_{i}$ be a set of continuous positive
functions on $M$, $U$ be a compact set $\subset M$ and $g$ the Lorentz metric
under study. We then define : $G(U,\epsilon_{i},g)=$ set of all Lorentz
metrics ${\widetilde g}$ such that :
$$
\left |{\partial^{i}{\widetilde g}\over \partial x^{i}}-
{\partial^{i}g\over \partial x^{i}} \right | <\epsilon_{i}
\; \; \; \; on \; U \; \forall i \; \; \; \; .
$$
In the compact-open topology, open sets are obtained from
the $G(U,\epsilon_{i},g)$ through the operations of arbitrary union and
finite intersection.
\vskip 0.3cm
\leftline {\bf 4.2.2. Open topology}
\vskip 0.3cm
We no longer require $U$ to be compact, and we take $U=M$ in section 4.2.1.
\vskip 0.3cm
\leftline {\bf 4.2.3. Fine topology}
\vskip 0.3cm
We define : $H(U,\epsilon_{i},g)=$ set of all Lorentz metrics $\widetilde g$
such that :
$$
\left | {\partial^{i}{\widetilde g}\over \partial x^{i}}-
{\partial ^{i}g \over \partial x^{i}} \right | <\epsilon_{i}
\; \; \; \; ,
$$
and ${\widetilde g}=g$ out of the compact set $U$. Moreover, we set :
$G'(\epsilon_{i},g)=\cup H(U,\epsilon_{i},g)$. A sub-basis for the fine
topology is then given by the neighbourhoods $G'(\epsilon_{i},g)$ [26].

Now, the underlying idea for stable causality is that space-time must not
contain closed timelike curves, and we still fail to find closed timelike
curves if we open out the null cones. In view of the former definitions,
this idea can be formulated as follows :
\vskip 0.3cm
\noindent
{\bf Definition 4.9.} A metric $g$ satisfies the stable causality condition
if, in the $C^{0}$ open
topology (see section 4.2.2.), an open neighbourhood of $g$ exists
no metric of which has closed timelike curves.

The Minkowski, FRW, Schwarzschild and Reissner-Nordstrom space-times are all
stably causal. If stable causality holds, the differentiable and conformal
structure can be determined from
the causal structure, and space-time cannot be compact
(because in a compact space-time there are closed timelike curves).
A very important characterization of stable causality is given
by the following theorem [12] :
\vskip 0.3cm
\noindent
{\bf Theorem 4.2.} A space-time $(M,g)$ is stably causal if and only if a
cosmic time function exists on $M$, namely a function whose gradient is
everywhere timelike.
\vskip 1cm
\leftline {\bf 4.3. Global hyperbolicity}
\vskip 1cm
Global hyperbolicity plays a key role in developing a rigorous theory of
geodesics in Lorentzian geometry and in proving singularity theorems. Its
ultimate meaning can be seen as requiring the existence of Cauchy surfaces,
namely spacelike hypersurfaces which each nonspacelike curve intersects
exactly once. In fact some authors [27] take this property as the
starting point in discussing global hyperbolicity. Indeed, Leray's original
idea was that the set of nonspacelike curves from $p$ to $q$ must be
compact in a suitable topology [28]. We shall here follow [12], [25] and [27]
defining and proving in part what follows.
\vskip 0.3cm
\noindent
{\bf Definition 4.10.} A space-time $(M,g)$ is globally hyperbolic if :

(a) strong causality holds ;

(b) $J^{+}(p)\cap J^{-}(q)$ is compact $\forall p,q \in M$.
\vskip 0.3cm
\noindent
{\bf Theorem 4.3.} In a globally hyperbolic space-time, the following
properties hold :

(1) $J^{+}(p)$ and $J^{-}(p)$ are closed $\forall p$ ;

(2) $\forall p,q$, the space $C(p,q)$ of all nonspacelike curves from $p$
to $q$ is compact in a suitable topology ;

(3) there are Cauchy surfaces.
\vskip 0.3cm
\noindent
{\bf Proof of (1).} It is well-known that, if $(X,F)$ is a Hausdorff
space and $A\subset X$ is compact, then $A$ is closed. In our case, this
implies that $J^{+}(p)\cap J^{-}(q)$ is closed. Moreover, it is not
difficult to see that $J^{+}(p)$ itself must be closed. In fact, otherwise
we could find a point $r \in {\overline {J^{+}(p)}}$ such that
$r\notin J^{+}(p)$. Let us now choose $q\in I^{+}(r)$. We would then have :
$r \in {\overline {J^{+}(p)\cap J^{-}(q)}}$ but
$r\notin J^{+}(p)\cap J^{-}(q)$, which implies that $J^{+}(p)\cap J^{-}(q)$
is not closed, not in agreement with what we found before. Similarly we also
prove that $J^{-}(p)$ is closed.

{\bf Remark} : a stronger result can also be proved. Namely, if $(M,g)$ is
globally hyperbolic and $K\subset M$ is compact, then $J^{+}(K)$ is closed
[27].
\vskip 0.3cm
\noindent
{\bf Proof of (3).} The proof will use the following ideas :
\vskip 0.3cm
\noindent
{\bf Step 1.} We define a function $f^{+}$, and we prove that global
hyperbolicity implies continuity of $f^{+}$ on $M$ [12].
\vskip 0.3cm
\noindent
{\bf Step 2.} We consider the function :
$$
f:p\in M \rightarrow f(p)\equiv {f^{-}(p)\over f^{+}(p)}
\; \; \; \; ,
\eqno (4.7)
$$
and we prove that the $f=constant$ surfaces are Cauchy surfaces [25].
\vskip 3cm
\centerline {\bf Step 1}
\vskip 0.3cm
The function $f^{+}$ we are looking for is given by
$f^{+} : p\in M \rightarrow$ volume of $J^{+}(p,M)$. This can only be
done with a suitable choice of measure. The measure is chosen in such a
way that the total volume of $M$ is equal to 1. For $f^{+}$ to be
continuous on $M$, it is sufficient
to show that $f^{+}$ is continuous on any
nonspacelike curve $\gamma$. In fact, let $r\in \gamma$, and let
$\{x_{n}\}$ be a sequence of points on $\gamma$ in the past of $r$. We now
define :
$$
T\equiv \cap J^{+}(x_{n},M) \; \; \; \; .
\eqno (4.8)
$$
If $f^{+}$ were not upper semi-continuous on $\gamma$ in $r$, there would
be a point $q\in T-J^{+}(r,M)$, with $r \notin J^{-}(q,M)$. But on the
other hand, the fact that $x_{n}\in J^{-}(q,M)$ implies that
$r\in {\overline {J^{-}(q,M)}}$, which is impossible in view of global
hyperbolicity. The absurd proves that $f^{+}$ is
upper semi-continuous. In the same way (exchanging the role of past and
future) we can prove lower semi-continuity, and thus continuity.
It becomes then trivial to prove the continuity of the function
$f^{+}: p \in M \rightarrow$ volume of $I^{+}(p,M)$. From now on, we shall
mean by $f^{+}$ the volume function of $I^{+}(p,M)$.
\vskip 0.3cm
\centerline {\bf Step 2}
\vskip 0.3cm
Let $\Sigma$ be the set of points where $f=1$, and let $p \in M$
be such that $f(p)>1$. The idea is to prove that every past-directed
timelike curve from $p$ intersects $\Sigma$, so that $p \in D^{+}(\Sigma)$.
In a similar way, if $f(p)<1$, one can then prove that $p \in D^{-}(\Sigma)$
(which finally implies that $\Sigma$ is indeed a Cauchy surface). The former
result can be proved as follows [25].
\vskip 3cm
\centerline {\bf Step 2a}
\vskip 0.3cm
We consider any past-directed timelike curve $\mu$ without past endpoint
from $p$. In view of the continuity of $f$ proved in step 1, such a
curve $\mu$ must intersect $\Sigma$, provided one can show that there is
$\epsilon \rightarrow 0^{+}$ : $f_{on \; \mu}=\epsilon$ , where
$\epsilon$ is arbitrary.
\vskip 0.3cm
\centerline {\bf Step 2b}
\vskip 0.3cm
Given $q \in M$, we denote by $U$ a subset of $M$ such that
$U \subset I^{+}(q)$. The subsets $U$ of this form cover $M$. Moreover,
any $U$ cannot be in $I^{-}(r)$ $\forall r$ $\in \mu$. This is forbidden by
global hyperbolicity. In fact, suppose for absurd that
$q \in \cap_{r \in \mu}I^{-}(r)$. We then choose a sequence $\{t_{i}\}$ of
points on $\mu$ such that :
$$
t_{i+1}\in I^{-}(t_{i}) \; \; \; \;
\exists i : z \in I^{-}(t_{i}) \; \forall z \in \mu
$$
$\forall i$, we also consider a timelike curve $\mu'$ such that :

(1) $\mu'$ begins at $p$ ;

(2) $\mu'=\mu$ to $t_{i}$ ;

(3) $\mu'$ continues to $q$.

Global hyperbolicity plays a role in ensuring that the sequence $\{t_{i}\}$
has a limit curve $\Omega$, which by construction contains $\mu$. On the
other hand, we know this is impossible. In fact, if $\mu$ were contained in
a causal curve from $p$ to $q$, it should have a past endpoint, which is
not in agreement with the hypothesis. Thus, having proved that
$\exists r \in \mu$ : $U \not \subset I^{-}(r)$, we find that
$f^{-}(r)\rightarrow 0$ when $r$ continues into the past on $\mu$, which
in turn implies that $\mu$ intersects $\Sigma$ as we said in step 2a [25].

The proof of (2) is not given here, and can be found in [12].
Global hyperbolicity plays a key role in proving singularity
theorems because, if $p$ and $q$ lie in a globally hyperbolic set and
$q \in J^{+}(p)$, there is a nonspacelike geodesic from $p$ to $q$
whose length is greater than or equal to that of any other nonspacelike
curve from $p$ to $q$. The proof that arbitrary, sufficiently small
variations in the metric do not destroy global hyperbolicity can be
found for example in [25]. Globally hyperbolic space-times are also
peculiar in that for them the Lorentzian distance function defined in
section 2.2.2. is finite and continuous as the Riemannian distance function
(see [1], p 86). The relation between strong causality,
finite distance function and global hyperbolicity is proved on p 107 of
[1]. More recent work on causal structure of Lorentzian
manifolds can be found in [29] and references therein.
\vskip 1cm
\leftline {\bf 5. ASYMPTOTIC STRUCTURE}
\vskip 1cm
Under this name one can discuss black holes theory, gravitational
radiation, positive mass theorems (for the ADM and Bondi's mass), the
singularity problem. Here we choose to focus on two topics : the asymptotic
symmetry group of space-time and the definition of boundary of space-time.
\vskip 1cm
\leftline {\bf 5.1. The Bondi-Metzner-Sachs group}
\vskip 1cm
For a generic space-time, the isometry group is simply the identity, and
thus does not provide relevant information. But isometry groups play a
very important role in physics. The most important example is given by the
Poincar\'e group, which is the group of all real transformations of
Minkowski space-time :
$$
x'=\Lambda x +a \; \; \; \; ,
\eqno (5.1)
$$
which leave invariant the length $(x-y)^{2}$. Namely, the Poincar\'e group
is given by the semidirect product
of the Lorentz group $O(3,1)$
and of translations $T_{4}$ in Minkowski space-time.

It is therefore very important to generalize the concept of isometry group
to a suitably regular curved space-time [3]. The
diffeomorphism group is not really useful because it is ''too large''
and it only preserves the differentiable structure of space-time. The
concept of asymptotic symmetry group makes sense for any space-time
$(M,g)$ which tends to infinity either to Minkowski or to a
Friedmann-Robertson-Walker model. The goal is achieved adding to $(M,g)$ a
boundary given by future null infinity, past null infinity or the whole of
null infinity (hereafter referred to as "scri").
We are now going to formulate in a precise way this idea.
For this purpose let us begin by recalling that the cuts of scri are
spacelike two-surfaces in scri orthogonal to the generators of scri. Each
cut has $S^2$ topology. They can be regarded as Riemann spheres with
coordinates $(\zeta,\zeta^{*})$, where $\zeta=x+iy$ and $\zeta^{*}$ is the
complex conjugate of $\zeta$, so that locally the metric is given by :
$ds^{2}=-d\zeta d\zeta^{*}$. Thus, defining [20] :
$$
\zeta \equiv e^{i\phi}\cot {\theta \over 2} \; \; \; \; ,
\eqno (5.2)
$$
we find :
$$
ds^{2}=-{1\over 4}(1+\zeta \zeta^{*})^{2}d\Sigma^{2} \; \; \; \; ,
\; \; \; \;
d\Sigma^{2}=d\theta^{2} +(\sin \theta)^{2}d\phi^{2} \; \; \; \; .
\eqno (5.3)
$$
Thus, if we choose a conformal factor $\Omega={2\over (1+\zeta \zeta^{*})}$,
each cut becomes the unit two-sphere. The choice of a chart can then be
used to define an asymptotic symmetry group. Indeed, the following simple but
fundamental result holds [20] :
\vskip 0.3cm
\noindent
{\bf Theorem 5.1.} All holomorphic bijections $f$ of the Riemann sphere are
of the form :
$$
\hat \zeta =f(\zeta)={{a\zeta +b}\over {c\zeta +d}} \; \; \; \; ,
\eqno (5.4)
$$
where $ad-bc=1$.

The transformations (5.4) are called fractional linear transformations
(FLT). Now, if a cut has to remain a unit sphere under (5.4), we must
perform another conformal transformation :
$d{\hat \Sigma}^{2}=K^{2}d\Sigma^{2}$, where [20] :
$$
K={{1+\zeta \zeta^{*}}\over
{(a\zeta +b)(a^{*}\zeta^{*}+b^{*})+(c\zeta +d)
(c^{*}\zeta^{*}+d^{*})}}
\; \; \; \; .
\eqno (5.5)
$$
Finally, for the theory to remain invariant under (5.5), the lengths along
the generators of scri must change according to : $d{\hat u}=Kdu$, which
implies :
$$
{\hat u}=K\Bigr [u+\alpha(\zeta,\zeta^{*}) \Bigr] \; \; \; \; .
\eqno (5.6)
$$
The transformations (5.4-6) form the Bondi-Metzner-Sachs (BMS)
asymptotic symmetry group of space-time. The subgroups of BMS are :
\vskip 0.3cm
\leftline {\bf 5.1.1. Supertranslations}
\vskip 0.3cm
This is the subgroup $S$ defined by :
$$
{\hat u}=u+\alpha(\zeta,\zeta^{*}) \; \; \; \; , \; \; \; \;
{\hat \zeta}=\zeta \; \; \; \; .
\eqno (5.7)
$$
The quotient group ${(BMS)\over S}$ represents the orthocronous proper
Lorentz group.
\vskip 0.3cm
\leftline {\bf 5.1.2. Translations}
\vskip 0.3cm
This four-parameter subgroup $T$ is given by (5.7) plus the following
relation :
$$
\alpha={{A+B\zeta +B^{*}\zeta^{*}+C\zeta \zeta^{*}}\over {1+\zeta \zeta^{*}}}
\; \; \; \; .
\eqno (5.8)
$$
The name is due to the fact that a translation in Minkowski space-time
generates a member of $T$. In fact, denoting by $(t,x,y,z)$ cartesian
coordinates in Minkowski space-time, if we set :
$$
u=t-r \; \; \; , \; \; \; r^{2}=x^{2}+y^{2}+z^{2}\; \; \; , \; \; \;
\zeta=e^{i\phi}\cot {\theta \over 2} \; \; \; , \; \; \;
Z={1\over {1+\zeta \zeta^{*}}} \; \; \; ,
\eqno (5.9)
$$
we find that [20] :
$$
Z^{2}\zeta=(x+iy){{1-{z\over r}}\over 4r} \; \; \; \; ,
\eqno (5.10)
$$
$$
x=r(\zeta +\zeta^{*})Z \; \; \; , \; \; \; y=-ir(\zeta -\zeta^{*})Z
\; \; \; , \; \; \;
z=r(\zeta \zeta^{*}-1)Z  \; \; \; .
\eqno (5.11)
$$
Thus the translation :
$$
t'=t+a \; \; \; , \; \; \; x'=x+b \; \; \; , \; \; \; y'=y+c \; \; \; ,
\; \; \; z'=z+d \; \; \; ,
\eqno (5.12)
$$
implies that :
$$
u'=u+Z\left(A+B\zeta +B^{*}\zeta^{*} +C\zeta \zeta^{*}\right)
+O \left({1\over r}\right) \; \; \; \; ,
\eqno (5.13)
$$
which agrees with (5.7-8).
\vskip 3cm
\leftline {\bf 5.1.3. Poincar\'e}
\vskip 0.3cm
A BMS transformation is obtained from a Lorentz transformation and a
supertranslation. This is why there are several Poincar\'e groups at scri,
one for each supertranslation which is not a translation, and no one of
them is preferred. This implies there is not yet agreement about how to
define angular momentum in an asymptotically flat space-time (because this
is related to the Lorentz group which is a part of the Poincar\'e group
as explained before). Still, the energy-momentum tensor is well-defined,
because it is only related to the translations.

The Poincar\'e group can be defined as the subgroup of BMS which maps
good cuts into good cuts [30]. Namely, there is a
four-parameter collection of cuts, called good cuts, whose asymptotic shear
vanishes. These good cuts provide the structure needed so as to reduce
BMS to the Poincar\'e group. In fact, the asymptotic shear
$\sigma^{0}(u,\zeta,\zeta^{*})$ of the $u=constant$ null surfaces is
related to the $(\sigma')^{0}(u',\zeta',(\zeta^{*})')$ of the
$u'=constant$ null surfaces through the relation :
$$
(\sigma')^{0}(u',\zeta',(\zeta^{*})')=K^{-1}
\Bigr[\sigma^{0}(u,\zeta,\zeta^{*})+ (edth)^{2}\alpha(\zeta,\zeta^{*})
\Bigr] \; \; \; \; ,
\eqno (5.14)
$$
where $\zeta'$ is the one given in (5.4), and the operator $edth$ is
defined on page 8 of [30]. In view of (5.7), for the
supertranslations the relation (5.14) assumes the form :
$$
(\sigma')^{0}(u,\zeta,\zeta^{*})=\sigma^{0}(u'-\alpha,\zeta,\zeta^{*})+
(edth)^{2}\alpha  \; \; \; \; .
\eqno (5.15)
$$
For stationary space-times (which have a timelike Killing vector field), the
Bondi system exists where $\sigma^{0}=0$. Therefore, a supertranslation
between two Bondi systems both having $\sigma^{0}=0$ leads to the equation :
$(edth)^{2}\alpha=0$, which is solved by the translation group. This
proves in turn that there is indeed a collection of good cuts as defined
before. As explained in [31] (see also [3]), in geometrical terms the main
ideas can be summarized as follows. The generators of scri are the integral
curves of a null vector field $N$.
A vector field $X$ is called an (asymptotic) symmetry if it generates
a diffeomorphism which leaves invariant the integral curves of $N$.
Denoting by $h$ the intrinsic metric on scri, one then has [31] :
$L_{X}N=-\rho N \; \; \; \; L_{X}h=2 \rho h \; \; \; \;
L_{N}\rho=0$ where $\rho$ is a smooth function. Any linear combination
and any Lie bracket of symmetries is still a symmetry, so that they form
a Lie algebra denoted by $B$, say. Given the vector field
$X= \beta N$, one finds that $X$ is a symmetry if and only if [31] :
$L_{N}\beta=0$. The symmetries of this form are the supertranslations
$ST \subset B$. As clarified in [31], $ST$ is the Abelian
infinite-dimensional
ideal of $B$, and the quotient $B/ST$ is found to be the Lie
algebra of the Lorentz group. As remarked in [32],
it should be emphasized that the basic problem in asymptotics,
namely the existence of solutions to Einstein's equations whose asymptotic
properties are described by the scri formalism, is still unsolved. We
refer the reader to [32] for a detailed study of this problem.
\vskip 1cm
\leftline {\bf 5.2. The boundary of space-time}
\vskip 1cm
The singularity theorems in general relativity [12]
were proved using a definition of singularities based on the $g$-boundary.
Namely, one defines a topological space, the $g$-boundary, whose points
are equivalence classes of incomplete nonspacelike geodesics. The points
of the $g$-boundary are then the singular points of space-time. As
emphasized for example in [8], this definition has two basic
drawbacks :

(1) it is based on geodesics, whereas in [33] it was proved there
are geodesically complete space-times with curves of finite length and
bounded acceleration ;

(2) there are several alternative ways of forming equivalence classes
and defining the topology.

Schmidt's method is along the following lines :
\vskip 0.3cm
\noindent
{\bf Step 1.} Connections are known to provide a parallelization of the
bundle $L(M)$ of linear frames.
\vskip 0.3cm
\noindent
{\bf Step 2.} This parallelization can be used to define a Riemannian metric.
\vskip 0.3cm
\noindent
{\bf Step 3.} This Riemannian metric has the effect of making
a connected component of
$L(M)$ into a metric space. This connected component $L'(M)$ is dense in
a complete metric space $L_{C}'(M)$.
\vskip 0.3cm
\noindent
{\bf Step 4.} One defines ${\overline M}$ as the set of orbits of the
transformation group on $L_{C}'(M)$.
\vskip 0.3cm
\noindent
{\bf Step 5.} The $b$-boundary $\partial M$ of $M$ is then defined as :
$\partial M \equiv {\overline M}-M$.
\vskip 0.3cm
\noindent
{\bf Step 6.} Singularities of $M$ are defined as points of the $b$-boundary
$\partial M$ which are contained in the $b$-boundary of any extension of
$M$.

A few more details about this construction can now be given.
\vskip 5cm
\centerline {\bf Step 1}
\vskip 0.3cm
The parallelization of $L(M)$ is obtained defining horizontal and
vertical vector fields. For this purpose, we denote at first by
$\pi : L(M)\rightarrow M$ the mapping of the frame at $x$ into $x$.
\vskip 0.3cm
\noindent
{\bf Definition 5.1.} The curve $\gamma$ in $L(M)$ is horizontal if the
frames $Y_{1}(t),...,Y_{n}(t)$ are parallel along $\pi(\gamma(t))$.
\vskip 0.3cm
\noindent
{\bf Definition 5.2.} The horizontal vector fields $B_{i}$ are the unique
vector fields such that :
$$
\pi_{*}((B_{i})_{\gamma})=Y_{i} \; \; \; \; , \; \; \; \;
\pi_{*}((B(\xi))_{\gamma})=\xi^{i}Y_{i} \; \; \; \; ,
\eqno (5.16)
$$
if $\gamma=Y_{1},...,Y_{n}$, where $\pi_{*}$ denotes as usual the
pull-back of $\pi$.
\vskip 0.3cm
\noindent
{\bf Definition 5.3.} Vertical vector fields are given by :
$$
\left(E^{*}\right)={\left({d\over dt}R_{a(t)\gamma}\right)}_{t=0}
\; \; \; \; ,
\eqno (5.17)
$$
where $R_{a}$ is the action of the general linear group $GL(n,R)$ on
$L(M)$. The parallelization of $L(M)$ is then given by
$\left(E_{k}^{* \; i},B_{i}\right)$.
\vskip 0.3cm
\centerline {\bf Step 2}
\vskip 0.3cm
\noindent
{\bf Definition 5.4.} Denoting by $gl(n,R)$ the Lie algebra of $GL(n,R)$,
a $gl(n,R)$-valued one-form $\omega$ is expressed as :
$$
\omega(Y)=\omega_{k}^{\; \; i}(Y)E_{i}^{\; \; k} \; \; \; \; .
\eqno (5.18)
$$
\vskip 0.3cm
\noindent
{\bf Definition 5.5.} The canonical 1-forms $\theta^{i}$ are given by :
$$
\pi_{*}(Y_{\gamma})=\theta^{i}(Y_{\gamma})Y_{i} \; \; \; \; ,
\eqno (5.19)
$$
if $\gamma=Y_{1},...,Y_{n}$.
\vskip 0.3cm
\noindent
{\bf Definition 5.6.} The Riemannian metric $g$ is then [8] :
$$
g(X,Y)=\sum_{i}\theta^{i}(X)\theta^{i}(Y)+\sum_{i,k}\omega_{k}^{\; \; i}(X)
\omega_{k}^{\; \; i}(Y) \; \; \; \; .
\eqno (5.20)
$$
\vskip 0.3cm
\centerline {\bf Step 3}
\vskip 0.3cm
The Riemannian metric $g$ defines a distance function according to (2.2).
Thus the connected component $L'(M)$ of $L(M)$ is a metric space, and it
uniquely determines a complete metric space, $L_{C}'(M)$. Moreover,
$L'(M)$ is dense in $L_{C}'(M)$.
\vskip 0.3cm
\centerline {\bf Step 4}
\vskip 0.3cm
One proves [8] that $GL(n,R)$ is a topological transformation
group on $L_{C}'(M)$, in that the transformations $R_{a}$ are uniformly
continuous and can be extended in a uniformly continuous way on the closure
of $L'(M)$ in $L_{C}'(M)$.

However, also Schmidt's definition has some drawbacks. In fact :

(1) in a closed FRW universe the initial and final singularities form the
same single point of the $b$-boundary [34] ;

(2) in the FRW and Schwarzschild solutions the $b$-boundary points are
not Hausdorff separated from the corresponding space-time [35].

A fully satisfactory improvement of Schmidt's definition is still an open
problem. Unfortunately, a recent attempt appeared in [9] was not correct.
\vskip 1cm
\leftline {\bf 6. HAMILTONIAN STRUCTURE}
\vskip 1cm
Dirac's theory of constrained Hamiltonian systems [36-37]
has been successfully applied to general relativity, though many unsolved
problems remain on quantization [10].
The ADM formalism for general relativity is discussed in [37-39].
The derivation of
boundary terms in the action integral can be found in [40-42],
whereas a modern treatment of the ADM
phase space for general relativity in the asymptotically flat case is in
[10]. More recently, Ashtekar's spinorial variables
have given rise to a renewed interest in canonical gravity [10].
We are now going to analyze
the "new" phase space of general relativity, only paying attention to the
classical theory.

The basic postulate of canonical gravity is that space-time is topologically
$\Sigma$x$R$, and it admits a foliation in spacelike three-manifolds
$\Sigma_{t}$, which are all diffeomorphic to $\Sigma$. Ashtekar's variables
for canonical gravity are very important at least for the following
reasons :

(1) they are one of the most striking applications of the spinorial
formalism to general relativity ;

(2) the constraint equations assume a polynomial form, which is not
achieved using the old variables ;

(3) they realize a formal analogy between gravity and Yang-Mills theory ;

(4) they could lead to an exact solution of the constraint equations of the
quantum theory.

The basic ideas of the formalism of $SU(2)$
spinors in Euclidean three-space are the following [10].
We consider $(V,\epsilon_{AB},G_{A'B})$, where
$V$ is a complex two-dimensional vector space with a nondegenerate
symplectic form $\epsilon_{AB}$ and a positive-definite Hermitian scalar
product $G_{A'B}$. Then, given a real three-manifold $\Sigma$, we take
the vector bundle $B$ over $\Sigma$ whose fibres are isomorphic to
$(V,\epsilon_{AB},G_{A'B})$. The $SU(2)$ spinor fields on $\Sigma$ are thus
the cross-sections of $B$. The isomorphism between the space of symmetric,
second-rank Hermitian spinors $\lambda^{AB}$ and the tangent space to
$\Sigma$ is realized by the soldering form $\sigma_{\; \; AB}^{a}$, and
the metric $h$ on $\Sigma$ is given by :
$$
h^{ab}\equiv \sigma_{\; \; AB}^{a}\sigma_{\; \; CD}^{b}
\epsilon^{AC}\epsilon^{BD}=-Tr{\left(\sigma^{a}\sigma^{b}\right)}
\; \; \; \; .
\eqno (6.1)
$$
The conjugation of $SU(2)$ spinors obeys the rules :
$$
{\left(\psi_{A}+\lambda \phi_{A} \right)}^{+}=\psi_{A}^{+} +
\lambda^{*}\phi_{A}^{+} \; \; \; \; , \; \; \; \;
{\left(\psi_{A}^{+}\right)}^{+}=-\psi_{A} \; \; \; \; ,
\eqno (6.2)
$$
$$
\epsilon_{AB}^{+}=\epsilon_{AB} \; \; \; \; , \; \; \; \;
{\left(\psi_{A}\phi_{B}\right)}^{+}=\psi_{A}^{+}\phi_{B}^{+} \; \; \; \; ,
\eqno (6.3)
$$
$$
{\left(\psi^{A}\right)}^{+}\psi_{A} >0
\; \; \; \; \; \; \; \; \forall \psi_{A} \not = 0 \; \; \; \; .
\eqno (6.4)
$$
We now
consider a new configuration space $C$ in the asymptotically flat case,
defined as the space of all $\sigma_{\; \; A}^{a \; \; \; B}$ such that [10] :
$$
\sigma_{\; \; A}^{a \; \; \; B}={\left[1+{M(\theta,\phi)\over r}\right]}^{2}
(\sigma^{0})_{\; \; A}^{a \; \; \; B}+
O\left({1\over r^{2}}\right) \; \; \; \; .
\eqno (6.5)
$$
The momentum conjugate to $\sigma_{\; \; A}^{a \; \; \; B}$, following
[10], is denoted by $M_{aA}^{\; \; \; \; \; B}$ and it obeys
the relations :
$$
Tr(M_{a}\sigma^{a})=O\left({1\over r^{3}}\right) \; \; \; \; ,
\eqno (6.6)
$$
$$
M_{aA}^{\; \; \; \; \; B}+{1\over 3}Tr\left(M_{l}\sigma^{l}\right)\sigma_{aA}
^{\; \; \; \; \; B}=O\left({1\over r^{2}}\right) \; \; \; \; .
\eqno (6.7)
$$
The extended phase space $\Gamma$ is the space whose points are the
pairs $\left(\sigma_{\; \; A}^{a \; \; \; B},M_{aA}^{\; \; \; \; \; B}
\right)$ obeying (6.5-7), and the Poisson brackets among observables
are defined by :
$$
\{u,v\}\equiv \int_{\Sigma}Tr \left(
{\delta u \over \delta M_{a}}{\delta v \over \delta \sigma^{a}}
-{\delta u \over \delta \sigma^{a}}{\delta v \over \delta M_{a}}
\right)\; d^{3}x \; \; \; \; .
\eqno (6.8)
$$
In going to the new phase space we have added three degrees of freedom, which
lead to three new constraints :
$$
C_{ab}=-Tr\left(M_{[a}\sigma_{b]}\right)=M_{[ab]} \; \; \; \; ,
\eqno (6.9)
$$
in addition to the Hamiltonian and momentum constraints. We are now going
to consider the phase space :
$$
\Gamma' \equiv \left \{\left({\widetilde \sigma}_{\; \; J}^{a \; \; \; L},
A_{aJ}^{\; \; \; \; \; L}\right)\right \} \; \; \; \; ,
\eqno (6.10)
$$
where the spinorial variables ${\widetilde \sigma}_{\; \; J}^{a \; \; \; L}$
and $A_{aJ}^{\; \; \; \; \; L}$ are obtained from
$\sigma_{\; \; J}^{a \; \; \; L}$ and $M_{aJ}^{\; \; \; \; \; L}$ as
follows. The variable ${\widetilde \sigma}_{\; \; J}^{a \; \; \; L}$
is defined by :
$$
{\widetilde \sigma}_{\; \; J}^{a \; \; \; L} \equiv
\sqrt{h}\sigma_{\; \; J}^{a \; \; \; L} \; \; \; \; .
\eqno (6.11)
$$
The step leading to $A_{aJ}^{\; \; \; \; \; L}$ is simple but not trivial
(it can be more thoroughly understood recalling the definition of Sen
connection as done in [10]). At first we define a new momentum
variable :
$$
\pi_{aJ}^{\; \; \; \; \; L}\equiv
{1\over \sqrt{h}}\left[
M_{aJ}^{\; \; \; \; \; L}+{1\over 2}Tr(M_{b}\sigma^{b})
\sigma_{aJ}^{\; \; \; \; \; L}
\right ] \; \; \; \; .
\eqno (6.12)
$$
Now, denoting by $D$ the connection on the real three-manifold $\Sigma$,
we define a new connection ${\widetilde D}$ by [10] :
$$
{\widetilde D}_{a}\lambda_{M}\equiv \partial_{a}\lambda_{M}+
A_{aM}^{\; \; \; \; \; C}\lambda_{C} \; \; \; \; .
\eqno (6.13)
$$
The spinorial variable $A_{aJ}^{\; \; \; \; \; L}$ in (6.13) is obtained
from (6.12) and from the spin-connection 1-form $\Gamma_{aJ}^{\; \; \; \; \;
L}$ of $D$ by :
$$
A_{aJ}^{\; \; \; \; \; L}\equiv \Gamma_{aJ}^{\; \; \; \; \; L}
+{i\over \sqrt{2}}\pi_{aJ}^{\; \; \; \; \; L} \; \; \; \; ,
\eqno (6.14)
$$
where the spin connection is known to be the unique connection which
annihilates the soldering form $\sigma_{\; \; JL}^{a}$, and is given by :
$$
\Gamma_{a}^{\; \; JL}\equiv -{1\over 2}\sigma_{f}^{\; \; EL}
\left[\partial_{a}\sigma_{\; \; \; \; E}^{fJ}+
\Gamma_{ba}^{f}\sigma_{\; \; \; \; E}^{bJ}\right] \; \; \; \; ,
\eqno (6.15)
$$
and $\Gamma_{ba}^{f}$ are the Christoffel symbols involving the three-metric
$h$ on $\Sigma$. The new variables defined in (6.11) and (6.14) obey
the Poisson bracket relations [10] :
$$
\Bigr \{{\widetilde \sigma}_{\; \; J}^{a \; \; \; L}(x),
{\widetilde \sigma}_{\; \; M}^{b \; \; \; N}(y)\Bigr \}=0 \; \; \; \; ,
\eqno (6.16)
$$
$$
\Bigr \{A_{a}^{\; \; JL}(x),{\widetilde \sigma}_{\; \; MN}^{m}(y)
\Bigr \}={i\over \sqrt{2}}\delta(x,y)\delta_{M}^{\; \; (J}
\delta_{N}^{\; \; L)} \; \; \; \; ,
\eqno (6.17)
$$
$$
\Bigr \{A_{aJ}^{\; \; \; \; \; L}(x),A_{bM}^{\; \; \; \; \; N}(y)
\Bigr \}=0 \; \; \; \; .
\eqno (6.18)
$$
Finally, denoting by $F_{abM}^{\; \; \; \; \; \; \; N}$ the curvature of
$\widetilde D$ :
$$
F_{abM}^{\; \; \; \; \; \; \; C}\lambda_{C}=
2{\widetilde D}_{[a}{\widetilde D}_{b]}\lambda_{M} \; \; \; \; ,
\eqno (6.19)
$$
the constraints of the theory assume the form [10] :
$$
{\widetilde D}_{a}\left({\widetilde \sigma}_{\; \; J}^{a \; \; \; L}
\right)=0 \; \; \; \; ,
\eqno (6.20)
$$
$$
Tr \left({\widetilde \sigma}^{a}F_{ab}\right)=0 \; \; \; \; ,
\eqno (6.21)
$$
$$
Tr \left({\widetilde \sigma}^{a}{\widetilde \sigma}^{b}F_{ab}\right)=0
\; \; \; \; ,
\eqno (6.22)
$$
where the Gauss-law constraint (6.20) is due to (6.9), and (6.21-22)
are respectively the momentum and Hamiltonian constraints. Setting :
$$
E^{a}\equiv {\widetilde \sigma}^{a} \; \; \; \; , \; \; \; \;
B^{a}\equiv {1\over 2}\epsilon^{abc}F_{bc} \; \; \; \; ,
\eqno (6.23)
$$
we see that the new phase space (6.10) can be thought as a submanifold
of the constrained phase space of a complexified $SU(2)$ Yang-Mills
theory (see remark (3) in the beginning of this section), and the
constraints are indeed polynomial as we anticipated. One can also
reverse things, and regard the $A_{aJ}^{\; \; \; \; \; L}$ as configuration
variables, so that their momentum conjugate becomes
${\widetilde \sigma}_{\; \; J}^{a \; \; \; L}$. In so doing the momentum
constraints remain linear and the Hamiltonian constraint remains quadratic
in the momenta [10].

Also, it should be emphasized that ${\widetilde \sigma}_{\; \; J}^{a \; \; \;
L}$ is real whereas $A_{aJ}^{\; \; \; \; \; L}$ is complex, so that they are
not conjugate variables in the usual sense. We can overcome this difficulty
going to the complex regime. Namely, we consider a complex phase space
$\Gamma_{C}$ whose points are defined on a real three-manifold $\Sigma$.
The real section $\Gamma$ of $\Gamma_{C}$ is then defined by [10] :
$$
{\left({\widetilde \sigma}^{a}\right)}^{+}=-{\widetilde \sigma}^{a}
\; \; \; \; ,
\eqno (6.24)
$$
$$
{\left(A_{aJ}^{\; \; \; \; \; L}-\Gamma_{aJ}^{\; \; \; \; \; L}\right)}^{+}
=-\left(A_{aJ}^{\; \; \; \; \; L}-\Gamma_{aJ}^{\; \; \; \; \; L}\right)
\; \; \; \; .
\eqno (6.25)
$$
In so doing we get back to (real Lorentzian) general relativity, whereas
real Euclidean general relativity is defined by the conditions where
(6.24) gets unchanged, whereas in (6.25) the spin-connection
1-form $\Gamma_{aJ}^{\; \; \; \; \; L}$ does not appear.

At the classical level, important work has been done in [43]. In that paper,
the author has shown that the trace of the extrinsic curvature tensor of
the boundary of space-time is the generating function for the canonical
transformation of the phase space of general relativity introduced by
Ashtekar. When torsion is nonvanishing, an additional boundary term is
present in the generating function, which has the effect of making the
action complex.
\vskip 1cm
\leftline {\bf 7. SINGULARITIES FOR THEORIES WITH TORSION}
\vskip 1cm
The singularity theorems of Penrose, Hawking and Geroch [12,44-49]
show that Einstein's general relativity leads to the occurrence of
singularities in cosmology in a rather generic way. On the other hand,
much work has also been done on alternative theories of gravitation
[50]. It
is by now well-known that when we describe gravity as the gauge theory of
the Poincar\'e group, this naturally leads to theories with torsion
[4-7]. The basic ideas can be summarized as
follows [7,51-52]. The holonomy theorems
imply that torsion and curvature are related respectively to the groups
of translations and of homogeneous transformations in the tangent vector
spaces to a manifold. The introduction of torsion related to spin gives
rise to a strong link between gravitation and particle physics, because it
extends the holonomy group to the translations. An enlightening discussion
of gauge translations can be found for example in [53-54].
In particular, the introduction of [54]
clarifies from the very beginning the main geometric role played by the
translations in the gauge group : they change a principal fibre bundle
having no special relationship between the points on the fibres and the
base manifold into the bundle of linear frames of the base manifold. When
we consider the gauge theory of the Poincar\'e group, we discover that the
gauge fields for the translation invariance are the orthonormal frames,
and the gauge field for Lorentz transformations is the part of the full
connection called contorsion [7]. From the point of view of fibre-bundle
theory, the possibility of defining torsion is a peculiarity of
relativistic theories of gravitation. Namely, the bundle $L(M)$ of linear
frames is soldered to the base $B=M$, whereas for gauge theories other
than gravitation the bundle $L(M)$ is loosely connected to $M$ [5].
Denoting by $\theta : TL(M) \rightarrow R^{4}$ the soldering form and by
$\omega$ a connection 1-form
on $L(M)$, the torsion 2-form $T$ is defined by [5] :
$T\equiv d\theta + \omega \wedge \theta$. The Poincar\'e group deserves
special consideration because it corresponds to an external symmetry, it
yields momentum and angular momentum conservation, and its translational
part can be seen as carrying matter through space-time [6].

At the very high densities present in the early universe, the effects of
spin can no longer be neglected [52].
Thus it is natural to address the
question : is there a rigorous theory of singularities in a space-time
with torsion ? The answer can only be found discussing at first the
properties of geodesics in a space-time with torsion, and trying to define
what is a singularity in such a theory.
\vskip 1cm
\leftline {\bf 7.1. Space-Times with torsion and their geodesics}
\vskip 1cm
A space-time with torsion (hereafter referred to as $U_{4}$ space-time) is
defined adding the following fourth requirement to the ones in section
2.1. :

(4) Given a linear $C^{r}$ connection ${\widetilde \nabla}$ which obeys
the metricity condition, a nonvanishing tensor :
$$
S(X,Y)={\widetilde \nabla}_{X}Y -{\widetilde \nabla}_{Y}X -[X,Y]
\; \; \; \; ,
\eqno (7.1)
$$
where $X$ and $Y$ are arbitrary $C^r$ vector fields and the square bracket
denotes their Lie bracket. The tensor ${S\over 2}$
is then called the torsion tensor (compare with [12]).

Now, it is well-known that the curve $\gamma$ is defined to be a
geodesic curve if its
tangent vector moves by parallel transport, so that $\nabla_{X}X$
is parallel to ${\left({\partial \over \partial t}\right)}_{\gamma}$
(see, however, comment before definition 7.1.).
A new parameter $s(t)$, called affine parameter, can always be found such
that, in local coordinates, this condition is finally expressed by the
equation :
$$
{d^{2}x^{a}\over ds^{2}}+\Gamma_{bc}^{\; \; \; a}
{dx^{b}\over ds}{dx^{c}\over ds}=0 \; \; \; \; .
\eqno (7.2)
$$
The geodesic equation (7.2) will now contain the effect of torsion through
the symmetric part $S_{\; \; (bc)}^{a}$
(not to be confused with the vanishing $S_{(bc)}^{\; \; \; \; \; a}$).
It is very useful to study this equation
in a case of cosmological interest. For example, in a closed FRW universe
the only nonvanishing components of the torsion tensor are the ones given
in [52], so that, setting $m=1,2,3$, (7.2) yields [11] :
$$
{d^{2}x^{0}\over ds^{2}}+a{da\over ds}{ds\over dt}c_{ii}
{\left({dx^{i}\over ds}\right)}^{2}=0 \; \; \; \; ,
\eqno (7.3)
$$
$$
{d^{2}x^{m}\over ds^{2}}+\Gamma_{ij}^{\; \; \; m}{dx^{i}\over ds}
{dx^{j}\over ds}
+2\left({1\over a}{da\over ds}{ds\over dt}-Q\right)
{dx^{0}\over ds}{dx^{m}\over ds}=0 \; \; \; \; .
\eqno (7.4)
$$
In (7.3), $c_{ii}$ are the diagonal components of the unit three-sphere
metric, and we are summing over all $i=1,2,3$. In (7.4), we used the
result of [52] according to which :
$$
\Gamma_{0m}^{\; \; \; \; m}={{\dot a}\over a}-2Q \; \; \; \; , \; \; \; \;
\Gamma_{m0}^{\; \; \; \; m}={{\dot a}\over a} \; \; \; \;
\forall m=1,2,3  \; \; \; \; .
\eqno (7.5)
$$
Of course, $\dot a$ denotes ${da\over ds}{ds\over dt}$. Now,
if the field equations are such that both
${1\over a}{da\over ds}{ds\over dt}$ and $Q$ remain finite for all values
of $s$, the model will be nonspacelike geodesically complete.
If a torsion singularity is thought as a point where torsion is infinite,
we are ruling out this possibility with our criterion, in addition to the
requirement that the scale factor never shrinks to zero.
Thus it seems that, whatever the physical
source of torsion is (spin or theories with quadratic Lagrangians etc.),
nonspacelike geodesic completeness is a concept of physical relevance
even though test particles do not move along geodesics [13].

An important comment is now in order. We have defined geodesics exactly
as important treatises do
in general relativity (see [1], p 403; [12], p 33)
for reasons which will become even more
clear studying maximal timelike geodesics in section 7.2.
However, our definition differs from the one adopted in [13].
In that paper, our geodesics are
just called autoparallel curves, whereas the authors interpret
as geodesics the curves of extremal length whose tangent
vector is parallelly transported according to the Christoffel connection.
In other words, if test particles were moving along extremal curves in
theories with torsion, there would be a strong reason for defining geodesics
and studying singularities only as done in [13]. But, as explained in [13],
the trajectories of particles differ from extremal curves and from the
curves we call geodesics. Thus it appears important to improve our
understanding, studying the mathematical properties of a singularity theory
based on the definition of geodesics as autoparallel curves. This definition,
involving the properties of the full connection, may be expected to have
physical relevance, imposing regularity conditions on the geometry
and on the torsion
of the cosmological model which is studied. Moreover,
in view of the fact that the definition of timelike, null and
spacelike vectors is not affected by the presence of torsion, the whole
theory of causal structure outlined in section 4 remains unchanged.
Combining this remark (also made in [13]) with the qualitative
argument concerning the geodesic equation, we here give the following
preliminary definition [11] :
\vskip 0.3cm
\noindent
{\bf Definition 7.1.} A $U_{4}$ space-time is singularity-free if it is
timelike and null
geodesically complete, where geodesics are defined as curves
whose tangent vector moves by parallel transport with respect to the full
$U_{4}$ connection.

This definition differs from the one given in [13] because we
rely on a different definition of geodesics, and it
has the drawbacks already illustrated in the beginning of
section 5.2. However, definition $7.1.$
is a preliminary definition which allows a direct comparison with
the corresponding situation in general relativity,
is generic in that it does not depend on the specific physical
theory which is the source of torsion and
it has physical relevance not only for a closed FRW model but also for
completely arbitrary models as we said before.
Thus we can now try to make the same (and eventually additional)
assumptions which lead to singularity theorems in general relativity, and
check whether one gets timelike and or null geodesic incompleteness.
Indeed, the extrinsic curvature tensor and the vorticity which appears in
the Raychaudhuri equation will now explicitly contain the effects of
torsion, and it is not a priori clear what is going
to happen. Namely, if one adopts the definition 7.1. as a preliminary
definition of singularities in a $U_{4}$ space-time, the main
issues to be studied seem to be [11] :

(1) How can we explain from first principles that a space-time which is
nonspacelike geodesically incomplete may become nonspacelike geodesically
complete in the presence of torsion ? And is the converse possible ?

(2) What happens in a $U_{4}$ space-time [13] under the
assumptions which lead
to the theorems of Penrose, Hawking and Geroch ?

We shall now partially study question (2) in the
next sub-section.
\vskip 1cm
\noindent
{\bf 7.2. A singularity theorem without causality assumptions
for $U_{4}$ space-times}
\vskip 1cm
In this section we shall denote by $R(X,Y)$ the four-dimensional
Ricci tensor with scalar curvature $R$, and by $K(X,Y)$ the extrinsic
curvature tensor of a spacelike three-surface. The energy-momentum tensor
will be written as $T(X,Y)$, so that the Einstein equations are :
$$
R(X,Y)-{1\over 2}g(X,Y)R=T(X,Y) \; \; \; \; .
\eqno (7.6)
$$
In so doing, we are absorbing the $8\pi G$ factor into the definition
of $T(X,Y)$.
For the case of general relativity, it was proved in [46] that
singularities must occur under certain assumptions, even though no
causality requirements are made. In fact, Hawking's result [12,46]
states that space-time cannot be timelike geodesically complete if :

(1) $R(X,X)\geq 0$ for any nonspacelike vector $X$ (which can also be
written in the form : $T(X,X)\geq g(X,X){T\over 2}$) ;

(2) there exists a compact spacelike three-surface $\Sigma$ without edge ;

(3) the trace $K$ of the extrinsic curvature tensor $K(X,Y)$ of $\Sigma$
is either everywhere positive or everywhere negative.

We are now going to study the following problem : is there a suitable
generalization of this theorem in the case of a $U_{4}$ space-time ?
Indeed, a careful examination of Hawking's proof (see [12],
p 273) shows that the arguments which should be modified
or adapted in a $U_{4}$
space-time are the ones involving the Raychaudhuri equation and the results
which prove the existence or the nonexistence of conjugate points. We
are now going to examine them in detail.
\vskip 0.3cm
\leftline {\bf 7.2.1. Raychaudhuri equation}
\vskip 0.3cm
The generalized Raychaudhuri equation in the ECSK theory of gravity has
been derived in [55-56] (see also [57-58]).
It turns out that, denoting by ${\widetilde \omega}_{ab}$
and $\sigma_{ab}$ respectively the vorticity and the shear tensors, the
expansion $\theta$ for a timelike congruence of curves obeys the
equation :
$$
{d\theta \over ds}=-(R(U,U)+2\sigma^{2}-2{\widetilde \omega}^{2})
-{\theta^{2}\over 3}
+{\widetilde \nabla}_{a}{\left(\dot U \right)}^{a} \; \; \; \; .
\eqno (7.7)
$$
In (7.7), $U$ is the unit timelike tangent vector, and we have set :
$$
2\sigma^{2}\equiv \sigma_{ab}\sigma^{ab} \; \; \; \; , \; \; \; \;
2{\widetilde \omega}^{2}\equiv
\left(\omega_{ab}+{1\over 2}{\widetilde S}_{ab}\right)
\left(\omega^{ab}+{1\over 2}{\widetilde S}^{ab}\right) \; \; \; \; ,
\eqno (7.8)
$$
where $\omega_{ab}$ is the vorticity tensor obtained from the Christoffel
symbols, and ${\widetilde S}_{bc}$ is obtained from the spin tensor
$\sigma_{bc}^{\; \; \; a}$ through a relation usually assumed to be of the
form [56,59] :
$$
\sigma_{bc}^{\; \; \; a}={\widetilde S}_{bc}U^{a} \; \; \; \; .
\eqno (7.9)
$$
\vskip 0.3cm
\leftline {\bf 7.2.2. Existence of conjugate points}
\vskip 0.3cm
Conjugate points are defined as in general relativity [12],
but bearing in mind that now the Riemann tensor
is the one obtained from the connection ${\widetilde \nabla}$ appearing
in (7.1) :
$$
R(X,Y,Z,W)=\Bigr[{\widetilde \nabla}_{X}{\widetilde \nabla}_{Y}g(W)
-{\widetilde \nabla}_{Y}{\widetilde \nabla}_{X}g(W)
-{\widetilde \nabla}_{[X,Y]}g(W)\Bigr](Z) \; \; \; \; .
\eqno (7.10)
$$
In general relativity, if one assumes that at $s_{0}$ one has
$\theta(s_{0})=\theta_{0}<0$, and $R(U,U)\geq 0$ everywhere, then one can
prove there is a point conjugate to $q$ along $\gamma(s)$ between
$\gamma(s_{0})$ and $\gamma \left(s_{0}-{3\over \theta_{0}}\right)$, provided
$\gamma(s)$ can be extended to $\gamma \left(s_{0}-{3\over
\theta_{0}}\right)$. This result is then extended to prove the existence
of points conjugate to a three-surface $\Sigma$ along $\gamma(s)$ within
a distance ${3\over \theta'}$ from $\Sigma$, where $\theta'$ is the initial
value of $\theta$ given by the trace $K$ of $K(X,Y)$, provided
$K<0$ and $\gamma(s)$ can be extended to that distance (see propositions
4.4.1 and 4.4.3 from [12]). This is achieved
studying an equation of the kind (7.7) where
${\widetilde \omega}^{2}=\omega^{2}$ is vanishing because $\omega_{ab}$
is constant and initially vanishing, and the last term on the right-hand
side vanishes as well. However, in the ECSK theory,
${\widetilde \omega}^{2}$ will still contribute in view of (7.8).
Thus the inequality :
$$
{d\theta \over ds}\leq -{\theta^{2}\over 3} \; \; \; \; ,
$$
can only make sense if we assume that :
$$
R(U,U)-2{\widetilde \omega}^{2}\geq 0 \; \; \; \; ,
\eqno (7.11)
$$
where we do not strictly need to include $2\sigma^{2}$ on the left-hand side
of (7.11) because $\sigma^{2}$ is positive [12,58].
If (7.11) holds, we can write (see (7.7) and set there
${\widetilde \nabla}_{a}{\left(\dot U \right)}^{a}=0$) :
$$
\int_{\theta_{0}}^{\theta}y^{-2}dy \leq -{1\over 3}\int_{s_{0}}^{s}dx
\; \; \; \; ,
\eqno (7.12)
$$
which implies :
$$
\theta \leq {3\over {s-\left(s_{0}-{3\over \theta_{0}}\right)}}
\; \; \; \; ,
\eqno (7.13)
$$
where $\theta_{0}<0$. Thus $\theta$ becomes infinite and there are
conjugate points for some $s\in \left]s_{0},s_{0}-{3\over \theta_{0}}\right]$.
However, (7.11) is a restriction on the torsion tensor. In fact, the
equations of the ECSK theory are given by (7.6) plus another one more
suitably written in the form used in [13] (compare with [11] and [59]) :
$$
S_{bc}^{\; \; \; a}-\delta_{b}^{a}S_{dc}^{\; \; \; d}
-\delta_{c}^{a}S_{bd}^{\; \; \; d}
=\sigma_{bc}^{\; \; \; a} \; \; \; \; .
\eqno (7.14)
$$
In (7.14)
we have absorbed the $8\pi G$ factor into the definition of
$\sigma_{bc}^{\; \; \; a}$ , whereas this is not done in (7.9). Setting
$\epsilon=g(U,U)=-1$, $\rho=8\pi G$, the insertion of
(7.9) into (7.14) and the multiplication by $U_{a}$ yields :
$$
{\widetilde S}_{bc}={1\over \rho \epsilon}
\left(U_{a}S_{bc}^{\; \; \; a}-U_{b}S_{dc}^{\; \; \; d}
-U_{c}S_{bd}^{\; \; \; d} \right) \; \; \; \; ,
\eqno (7.15)
$$
which implies, defining :
$$ \eqalignno{
f(\omega,\omega S) &\equiv \omega_{ab}\omega^{ab}+{1\over 2}
\omega_{ab}{\widetilde S}^{ab}+{1\over 2}{\widetilde S}_{ab}
\omega^{ab}\cr
&=\omega_{ab}\omega^{ab}+{\omega_{ab}\over 2\rho \epsilon}
\left(U_{h}S^{abh}-U^{a}S_{h}^{\; \; bh}
-U^{b}S_{\; \; h}^{a \; \; h}\right)\cr
&+{\omega^{ab}\over 2\rho \epsilon}\left(U_{h}S_{ab}^{\; \; \; h}
-U_{a}S_{hb}^{\; \; \; h}-U_{b}S_{ah}^{\; \; \; h}\right) \; \; \; \; ,
&(7.16)\cr}
$$
and using (7.8) and (7.11), that :
$$ \eqalignno{
8{\widetilde \omega}^{2}&=4f(\omega,\omega S)+{\widetilde S}_{bc}
{\widetilde S}^{bc}\cr
&=4f(\omega,\omega S)+
\rho^{-2} \Bigr(U_{h}S_{bc}^{\; \; \; h}-U_{b}S_{hc}^{\; \; \; h}
-U_{c}S_{bh}^{\; \; \; h}\Bigr)
\Bigr(U_{f}S^{bcf}-U^{b}S_{f}^{\; \; cf}-U^{c}S_{\; \; f}^{b \; \; f}
\Bigr)\cr
&\leq 4 R(U,U) \; \; \; \; .
&(7.17)\cr}
$$
Indeed some cases have been studied [55]
where $\omega_{ab}$ is vanishing.
However we here prefer to write the equations in general form.
Moreover, in extending (7.13) so as to prove the existence of conjugate
points to spacelike three-surfaces, the assumption $K<0$ on the trace
$K$ of $K(X,Y)$ also implies another condition on the torsion tensor. In
fact, denoting by $\chi(X,Y)$ the tensor obtained from the metric and from
the lapse and shift functions as the extrinsic curvature in general
relativity, in a $U_{4}$ space-time one has :
$$
K(X,Y)=\chi(X,Y)+\lambda(X,Y) \; \; \; \; ,
\eqno (7.18)
$$
where the symmetric part of $\lambda(X,Y)$ (the only one which contributes
to $K$) is given by :
$$
\lambda_{(ab)}=-2n^{\mu}S_{(a\mu b)} \; \; \; \; .
\eqno (7.19)
$$
In (7.19) we have changed sign with respect to Pilati [60] because his
convention for $K(X,Y)$ is opposite to Hawking's convention, and we are
here following Hawking so as to avoid confusion in comparing theorems.
Thus the condition $K<0$ implies the following restriction on torsion :
$$
\lambda=-2g^{ab}n^{\mu}S_{(a\mu b)}<-\chi \; \; \; \; .
\eqno (7.20)
$$
When (7.11) and (7.20) hold, one follows exactly the same technique
which leads to (7.13) in proving there are points conjugate to a
spacelike three-surface.
\vskip 0.3cm
\leftline {\bf 7.2.3. Maximal timelike geodesics}
\vskip 0.3cm
In general relativity, it is known (proposition 4.5.8 of [12])
that a timelike geodesic curve $\gamma$ from $q$ to $p$ is
maximal if and only if there is no point conjugate to $q$ along
$\gamma$ in $(q,p)$. We are now
going to sum up how this result is proved and then extended so as to rule out
the existence of points conjugate to three-surfaces. This last step will then
be enlightening in understanding what changes in a $U_{4}$ space-time [11].

We shall here follow the conventions of section 4.5 of [12],
denoting by $L(Z_{1},Z_{2})$ the second derivative of the arc length
defined in (2.3), by $V$ the unit tangent vector ${\partial \over
\partial s}$ and by $T_{\gamma}$ the vector space consisting of all
continuous, piecewise $C^2$ vector fields along the timelike geodesic
$\gamma$ orthogonal to $V$ and vanishing at $q$ and $p$. We are here just
interested in proving that, if the timelike geodesic $\gamma$ from $q$
to $p$ is maximal, this implies there is no point conjugate to $q$. The
idea is to suppose for absurd that $\gamma$ is maximal but there is a
point conjugate to $q$. One then finds that $L(Z,Z)>0$, which in turn
implies that $\gamma$ is not maximal, against the hypothesis. This is achieved
taking a Jacobi field $W$ along $\gamma$ vanishing at $q$ and $r$, and
extending it to $p$ putting $W=0$ in the interval $[r,p]$. Moreover, one
considers a vector $M \in T_{\gamma}$ so that $g\left(M,{D\over \partial s}W
\right)=-1$ at $r$. In what follows, we shall just say that $M$ is suitably
chosen, in a way which will become clear later. One then defines :
$$
Z \equiv \epsilon M +\epsilon^{-1}W \; \; \; \; ,
\eqno (7.21)
$$
where $\epsilon$ is positive and constant. Thus, the general formula for
$L(Z_{1},Z_{2})$
implies that (see lemma 4.5.6 of [12]) :
$$
L(Z,Z)=\epsilon^{2}L(M,M)+2L(W,M)+\epsilon^{-2}L(W,W)
=\epsilon^{2}L(M,M)+2 \; \; \; \; ,
\eqno (7.22)
$$
which implies that $L(Z,Z)$ is $>0$ if $\epsilon$ is suitably small, as
we anticipated. The same method is also used in proving there cannot be
points conjugate to a three-surface $\Sigma$ if the timelike geodesic
$\gamma$ from $\Sigma$ to $p$ is maximal. However, as proved in lemma
4.5.7 of [12], in the case of a three-surface $\Sigma$,
the formula for $L(Z_{1},Z_{2})$ is of the kind :
$$
L(Z_{1},Z_{2})=F(Z_{1},Z_{2})-\chi(Z_{1},Z_{2}) \; \; \; \; ,
\eqno (7.23)
$$
where $\chi(X,Y)$ is the extrinsic curvature tensor of $\Sigma$. But we
know that in a $U_{4}$ space-time $\chi(X,Y)$ gets replaced by the
nonsymmetric tensor $K(X,Y)$ defined in (7.18-19), which can be
completed with the relation for the antisymmetric part of $\lambda(X,Y)$ :
$$
\lambda_{[ab]}=-n^{\mu}S_{ba \mu} \; \; \; \; .
\eqno (7.24)
$$
Thus now the splitting (7.21) leads to a formula of the kind (7.22)
where the requirement
$$
L(W,M)+L(M,W)=c>0 \; \; \; \; ,
\eqno (7.25)
$$
will involve torsion because (7.23) gets replaced by :
$$
L(Z_{1},Z_{2})={\widetilde F}(Z_{1},Z_{2})-K(Z_{1},Z_{2}) \; \; \; \; .
\eqno (7.26)
$$
Namely, the left-hand side of (7.25) will contain $K(W,M)+K(M,W)$.
The condition (7.25) also clarifies how to suitably choose $M$ in a
$U_{4}$ space-time. It is worth emphasizing that only $\lambda_{(ab)}$
contributes to (7.25) because the contributions of $\lambda_{[ab]}$
coming from $K(M,W)$ and $K(W,M)$ add up to zero. In proving (7.26),
the first step is the generalization of lemma 4.5.4 of [12]
to a $U_{4}$ space-time. This is achieved remarking that the
relation :
$$
{\partial \over \partial u}g \left({\partial \over \partial t},
{\partial \over \partial t} \right)=
2g \left({D\over \partial u}{\partial \over \partial t},
{\partial \over \partial t}\right) \; \; \; \; ,
\eqno (7.27)
$$
is also valid in a $U_{4}$ space-time, where now ${D\over \partial u}$
denotes the covariant derivative along the curve with respect to the full
$U_{4}$ connection. In fact, denoting by $X$ the vector ${\partial \over
\partial t}$ and using the definition of covariant derivative along a curve
one finds [11] :
$$
{\partial \over \partial u}g(X,X)=2g \left({D\over \partial u}X,X \right)
+X^{a}X^{b}{D\over \partial u}g_{ab} \; \; \; \; ,
\eqno (7.28)
$$
where ${D\over \partial u}g_{ab}$ is vanishing if the connection obeys the
metricity condition, which is also assumed in a $U_{4}$ space-time
(see section 7.1. and [4]). In other words, the key role is
played by the connection which obeys the metricity condition, and
${\partial \over \partial u}g(X,X)$ will implicitly contain the effects of
torsion because of the relation :
$$
{DX^{a}\over \partial u}\equiv {\partial X^{a}\over \partial u}
+\Gamma_{bc}^{\; \; \; a}{dx^{b}\over du}X^{c} \; \; \; \; .
\eqno (7.29)
$$
Although this point seems to be elementary, it plays a vital role in
leading to (7.26). This is why we chose to emphasize it [11].
\vskip 0.3cm
\leftline {\bf 7.2.4. The singularity theorem}
\vskip 0.3cm
If we now compare the results discussed or proved in
sections 7.2.1-3. with p
273 of [12], we are led to state the following singularity theorem :
\vskip 0.3cm
\noindent
{\bf Theorem 7.1.} The $U_{4}$ space-time of the ECSK theory cannot be
timelike geodesically complete if :

(1) $R(U,U)-2{\widetilde \omega}^{2} \geq 0$ for any
nonspacelike vector $U$ ;

(2) there exists a compact spacelike three-surface $S$ without edge ;

(3) the trace $K$ of the extrinsic curvature tensor $K(X,Y)$ of $S$ is
either everywhere positive or everywhere negative, and
$L(W,M)+L(M,W)=c>0$ as defined in (7.25-26) and before.

Conditions (1) and (3) will then involve the torsion tensor
defined through (7.1). Indeed, the second part of condition (3)
can be seen as a
prerequisite, but we have chosen to insert it into the statement of the
theorem so as to present together all conditions which involve the
extrinsic curvature tensor $K(X,Y)$.
The compatibility of (1) with the field
equations of the ECSK theory is expressed by (7.17)
whenever (7.9) makes sense. Otherwise, (7.17) should be replaced by a
different relation.
It is worth emphasizing that if we switch off torsion, condition (1)
becomes the one required in general relativity because, as
explained on pp 96-97 of [12], the vorticity of the
torsion-free connection vanishes wherever a 3x3 matrix which appears in the
Jacobi fields is nonsingular. Finally, if ${\widetilde \nabla}_{a}
{\left(\dot U \right)}^{a}$ is not vanishing as we assumed so far (see
(7.7) and comment before (7.12)) following [55-56],
the condition (1) of our theorem should be replaced by [11] :

($1'$) $R(U,U)-2{\widetilde \omega}^{2}-{\widetilde \nabla}_{a}
{\left(\dot U \right)}^{a} \geq 0$ for any nonspacelike vector $U$.
\vskip 1cm
\leftline {\bf 8. CONCLUDING REMARKS}
\vskip 1cm
At first we have seen our task as presenting an unified description of some
aspects of the differentiable, spinor, causal, asymptotic and Hamiltonian
structure of space-time. There is a very rich literature on these topics
on specialized books [1,3,10,12,14,16-18] and on the original papers
(see also [61-78]), but we thought it was important to analyze them in a
single paper. This choice of arguments helps also nonexpert readers in
gaining familiarity with concepts and techniques frequently used in
classical gravity and which also find application in quantum gravity, as it
happens for the theory of spinors and for Ashtekar's variables.
Moreover, causal and asymptotic structure play a key role for singularity
theory in cosmology, which is a basic problem of classical theories of
gravity and constitutes the main motivation for the study of quantum
cosmology (see papers in [79-80]).
As a partial completion of what we studied so far, the following remarks
and mentions are now in order.

(a) An alternative description of space-time has been recently proposed in
[81]. In that paper, the author has shown that the gravitational field in
general relativity has the properties of a parametric manifold, namely a
mathematical structure generalizing the concept of gauge fields. The author
then explains how space-time can be seen as a parametric three-manifold
supplied with a metric and a gravitational potential, and he develops the
theory of parametric spinors [81].

(b) Relativists are quite often interested in four-dimensional space-time
models with the associated two-component spinor language [21].
However, from the point of view of theoretical physics, different
formalisms are also of interest. Thus the readers can look at [82-83],
and also consider a classical paper
such as [84]. Other remarkable results on spinors are the
ones proved in [85].

(c) In section 4.2. we briefly motivated and described stable causality.
Recent progress on the topology of stable causality is due to [86].
The authors give an enlightening
discussion of causally convex and stably causally convex sets and of their
topologies. For example, they prove that a point of space-time has
arbitrarily small neighbourhoods that are stably causally convex sets if
and only if stable causality holds. They also define returning sets,
analyzing the structure of subsets that control the fulfillment of
strong and stable causality at a point. In [87] volume functions have been
used to characterize strong causality, global hyperbolicity and other
causality conditions, and in [88] the causal boundary for stably causal
space-times has been analyzed. In [89], strong and stable causality have
been characterized in terms of causal functions through two important
theorems.

(d) There is a very rich literature on the asymptotic structure of
space-time [90-91].
For example, the structure of the gravitational field at
spatial infinity is studied through an analysis of asymptotically
Euclidean and Minkowskian space-times in [92-94].
More recently, impressive progress has been made in studying the
global structure of simple space-times in [95]. In that paper,
the author has proved what follows [95] :

(1) Future null infinity is diffeomorphic to the complement of a point in
some contractible open three-manifold ;

(2) The strongly causal region $\Sigma_{SC}$ of future null infinity is
diffeomorphic to $S^{2}$x$R$ ;

(3) Every compact connected spacelike two-surface in future null infinity
is contained in $\Sigma_{SC}$ ;

(4) Space-Time must be globally hyperbolic.

(e) In section 6 we only focused on some classical aspects of Ashtekar's
formalism for canonical gravity. However its main motivation is the
development of a nonperturbative approach to quantum gravity. At
present, the main interest is in a representation where quantum states
arise as functions of loops on a three-manifold, and in so doing a class
of exact solutions to all quantum constraints has been obtained for the
first time (see [96] and references therein).

(f) Very recent progress in singularity theory in cosmology is due to
[97-98]. In [97] the author has studied an
alternative interpretation according to which gravitational collapse
may give rise not to singularities but to chronology violation. He has
found an example of a singularity-free chronology violating space-time
with a nonachronal closed trapped surface. In [98], a remarkable
proof has been given that a nonempty chronology violation set with
compact boundary causes singularities. In other words, these papers shed
new light on the problem of whether causality violations lead to the
occurrence of singularities, and one can now prove that causality violations
that do not extend to infinity must cause singularities (see [98]
and references therein). In [99] it has been shown that a large class of
time-dependent solutions to Einstein's equations are classical solutions
to string theory. Interestingly enough, these include metrics with large
curvature and some with space-time singularities.

Finally, in section 7 we have studied other aspects of the singularity
problem in cosmology. We have then
taken the point of view according to which nonspacelike
geodesic incompleteness can be used as a preliminary definition of
singularities also in space-times with torsion. We have finally been able
to show under which conditions Hawking's singularity theorem without
causality assumptions can be extended to the space-time of the ECSK theory.
However, when we assume (7.9) and
we require consistency of the additional condition (7.11)
with the equations of the ECSK theory, we end up with the relation (7.17)
which explicitly involves the torsion tensor on the left-hand side
(of course, the torsion tensor is also present in $R(U,U)$ through the
connection coefficients, but this is an implicit appearance of torsion,
and it is better not to make this splitting). Also the conditions (7.20)
and (7.25) involve the torsion tensor in an explicit way
if one uses the formula (7.18).
This is why we interpret our result as an
indication of the fact that the presence of singularities in the ECSK theory
is less generic than in general relativity.
Our result should be compared with [13]. The relevant
differences between our work and their work are :

(1) We rely on a different definition of geodesics, as explained in
section 7.1. ;

(2) We emphasize the role played by the full extrinsic curvature tensor
and by the variation formulae in
$U_{4}$ theory, a remark which is absent in [13] ;

(3) We keep the field equations of the ECSK theory in their original form,
whereas the authors in [13] cast them in a form analogous to general
relativity, but with a modified energy-momentum tensor which contains
torsion. We think this technique is not strictly needed [11]. Moreover,
from a Hamiltonian point of view, the splitting of the Riemann tensor into
the one obtained from the Christoffel symbols plus the one explicitly
related to torsion does not seem to be in agreement with the choice of the
full connection as a canonical variable. In fact, if we look
for example at models with quadratic Lagrangians in $U_{4}$ theory, the frame
and the full connection should be regarded as independent variables
[52], and this choice of canonical variables has also been
made for the ECSK theory [100-102].

Problems to be studied for further research are the generalization to
$U_{4}$ space-times of the other singularity theorems in [12]
using our approach,
and of the results in [8] that we outlined in section 5.2.
Moreover, the generalization to $U_{4}$ space-times of the classification
of singularities appeared in [69], and its relation to the preliminary
definition of singularities we used in this paper (namely specification
of the regularity condition needed for the Riemann tensor and for the full
connection) deserves careful consideration.

Recently, the singularity problem for space-times with torsion has also been
studied in [103] for the case of classical $N=1$ supergravity. In that
paper the author has used the modified weak-energy condition for theories
with torsion developed in [13]. Pullin has found that spin-spin contact
interactions cannot avert singularities in general.
Finally, he has presented a
singularity-free model for a spatially homogeneous Rarita-Schwinger
field in a FRW space-time. Another recent approach to the gauge theory of
gravity is the one in [104], where the authors have given a
coordinate-free description of the $SO(3,2)$ theory of gravity. The groups
$SO(3,2)$ and $SO(4,1)$ are of special interest because they are the only
ones reducing to the Poincar\'e group by a process called
Wigner-Inonu contraction [105]. $SO(3,2)$ is the group which leads to
supersymmetry as a natural extension. The geometric analysis in [104]
yields a better understanding of the embedding of vierbein and Lorentz
connection into the connection of a larger symmetry group.
\vskip 1cm
\leftline {\bf Acknowledgements}
\vskip 1cm
We here wish to express our gratitude to Peter D'Eath and Stephen Hawking
for stimulating and encouraging our work.
We are also very much indebted to John Beem, Paolo Scudellaro, John Stewart
and Cosimo Stornaiolo for correspondence and conversations.
\vskip 40cm
\leftline {\bf REFERENCES}
\vskip 1cm
\item {[1]}
J. K. BEEM, P. E. EHRLICH, {\it Global Lorentzian Geometry} (Dekker,
New York, 1981).
\item {[2]}
P. BUDINICH, A. TRAUTMAN,
J. Geom. Phys. {\bf 4} (1987) 361.
\item {[3]}
R. PENROSE, W. RINDLER, {\it Spinors and Space-Time}, Vol. I, II.
(Cambridge Univ. Pres, Cambridge, 1984 and 1986).
\item {[4]}
F. W. HEHL, P. VON DER HEYDE, G. D. KERLICK, J. M. NESTER, Rev. Mod.
Phys. {\bf 48} (1976) 393.
\item {[5]}
A. TRAUTMAN, in
General Relativity and Gravitation, Vol. I, ed. A. Held (Plenum Press,
New York, 1980) 287.
\item {[6]}
F. W. HEHL, J. NITSCH, P. VON DER HEYDE, in General
Relativity and Gravitation, Vol. I, ed. A. Held (Plenum Press, New York,
1980) 329.
\item {[7]}
J. M. NESTER, in An
Introduction to Kaluza-Klein Theories, ed. H. C. Lee
(World Scientific, Singapore, 1983) 83.
\item {[8]}
B. G. SCHMIDT,
Gen. Rel. Grav. {\bf 1} (1971) 269.
\item {[9]}
J. GRUSZCZAK, M. HELLER, Z. POGODA, {\it A Singular Boundary of the Closed
Friedmann Universe} (1989), Cracow preprint, TPJU 11 (unpublished).
\item {[10]}
A. ASHTEKAR, {\it New Perspectives in Canonical Gravity} (Bibliopolis,
Napoli, 1988).
\item {[11]}
G. ESPOSITO,
Nuovo Cimento {\bf B 105} (1990) 75.
\item {[12]}
S. W. HAWKING, G. F. R. ELLIS, {\it The Large Scale Structure of
Space-Time} (Cambridge Univ. Press, Cambridge, 1973).
\item {[13]}
F. W. HEHL, P. VON DER HEYDE, G. D. KERLICK, Phys. Rev.
{\bf D 10} (1974) 1066.
\item {[14]}
R. PENROSE, {\it Techniques of Differential Topology in Relativity}
(Society for Industrial and Applied Mathematics, Bristol, 1983).
\item {[15]}
S. W. HAWKING, in
General Relativity, an Einstein Centenary Survey, eds. S. W.
Hawking, W. Israel (Cambridge Univ. Press, Cambridge, 1979) 746.
\item {[16]}
S. GALLOT, D. HULIN, J. LAFONTAINE, {\it Riemannian Geometry}
(Springer-Verlag, Berlin, 1987).
\item {[17]}
M. FRANCAVIGLIA, {\it Elements of Differential and Riemannian Geometry}
(Bibliopolis, Napoli, 1988).
\item {[18]}
B. O'NEILL, {\it Semi-Riemannian Geometry with Applications to Relativity},
(Academic Press, New York, 1983).
\item {[19]}
R. PENROSE, in Battelle Rencontres (Benjamin, New York, 1968) 121.
\item {[20]}
J. M. STEWART, {\it Part III Lecture Notes in Applications of General
Relativity} (University of Cambridge, 1985).
\item {[21]}
Z. PERJ\'ES,
Proc. Centre Math. Anal. Austral. Nat. Univ. {\bf 19} (1989) 207.
\item {[22]}
R. S. WARD,
Commun. Math. Phys. {\bf 78} (1980) 1.
\item {[23]}
R. GEROCH,
J. Math. Phys. {\bf 9} (1968) 1739.
\item {[24]}
J. W. MILNOR, J. D. STASHEFF, {\it Characteristic Classes} (Princeton
University Press, Princeton, 1974).
\item {[25]}
R. GEROCH, J. Math. Phys. {\bf 11} (1970) 437.
\item {[26]}
S. W. HAWKING,
Gen. Rel. Grav. {\bf 1} (1971) 393.
\item {[27]}
R. M. WALD, {\it General Relativity} (The University of Chicago Press,
Chicago, 1984).
\item {[28]}
J. LERAY, {\it Hyperbolic Partial Differential Equations} (Princeton, 1952).
\item {[29]}
A. LEVICHEV,
Gen. Rel. Grav. {\bf 21} (1989) 1027.
\item {[30]}
E. T. NEWMAN, K. P. TOD, in
General Relativity and Gravitation, Vol. II, ed. A. Held
(Plenum Press, New York, 1980) 1.
\item {[31]}
A. ASHTEKAR, {\it Asymptotic Quantization} (Bibliopolis, Napoli, 1987).
\item {[32]}
B. G. SCHMIDT, J. M. STEWART,
Proc. R. Soc. Lond. {\bf A 420} (1988) 355.
\item {[33]}
R. GEROCH,
Ann. of Phys. {\bf 48} (1968) 526.
\item {[34]}
B. BOSSHARD,
Commun. Math. Phys. {\bf 46} (1976) 263.
\item {[35]}
R. A. JOHNSON, J. Math. Phys. {\bf 18} (1977) 898.
\item {[36]}
P. A. M. DIRAC, {\it Lectures on Quantum Mechanics}, Belfare Graduate
School of Science (Yeshiva University, New York, 1964).
\item {[37]}
T. REGGE, A. HANSON, C. TEITELBOIM, {\it Constrained Hamiltonian
Systems} (Accademia Nazionale dei Lincei, Roma, 1976).
\item {[38]}
C. W. MISNER, K. S. THORNE, J. A. WHEELER, {\it Gravitation} (Freeman,
S. Francisco, 1973).
\item {[39]}
M. A. H. MAC CALLUM, in Quantum
gravity : an Oxford symposium, eds. C. J. Isham, R. Penrose,
D. W. Sciama (Clarendon Press, Oxford, 1975) 174.
\item {[40]}
J. W. YORK,
Phys. Rev. Lett. {\bf 28} (1972) 1082.
\item {[41]}
G. W. GIBBONS, S. W. HAWKING,
Phys. Rev. {\bf D 15} (1977) 2752.
\item {[42]}
J. W. YORK,
Found. of Phys. {\bf 16} (1986) 249.
\item {[43]}
B. P. DOLAN, Phys. Lett. {\bf 233 B} (1989) 89.
\item {[44]}
S. W. HAWKING,
Proc. R. Soc. Lond. {\bf A 294} (1966) 511.
\item {[45]}
S. W. HAWKING,
Proc. R. Soc. Lond. {\bf A 295} (1966) 490.
\item {[46]}
S. W. HAWKING,
Proc. R. Soc. Lond. {\bf A 300} (1967) 187.
\item {[47]}
R. GEROCH, Phys. Rev. Lett. {\bf 17} (1966) 445.
\item {[48]}
S. W. HAWKING, R. PENROSE,
Proc. R. Soc. Lond. {\bf A 314} (1970) 529.
\item {[49]}
R. GEROCH, G. T. HOROWITZ, in
General Relativity, an Einstein Centenary Survey, eds. S. W.
Hawking, W. Israel (Cambridge Univ. Press, Cambridge, 1979) 212.
\item {[50]}
H. FUCHS, $\; \;$V. KASPER, D.-E. LIEBSCHER, V. MULLER, H.-J. SCHMIDT,
Fortschr. Phys. {\bf 36} (1988) 427.
\item {[51]}
A. TRAUTMAN,
Symposia Mathematica {\bf 12} (1973) 139.
\item {[52]}
G. ESPOSITO,
Nuovo Cimento B {\bf 104} (1989) 199.
\item {[53]}
E. A. LORD, P. GOSWAMI,
J. Math. Phys. {\bf 29} (1988) 258.
\item {[54]}
P. K. SMRZ,
J. Math. Phys. {\bf 28} (1987) 2824.
\item {[55]}
J. M. STEWART, P. H\'AJICEK,
Nat. Phys. Sci. {\bf 244} (1973) 96.
\item {[56]}
J. TAFEL,
Phys. Lett. {\bf 45 A} (1973) 341.
\item {[57]}
A. K. RAYCHAUDHURI,
Phys. Rev. {\bf D 12} (1975) 952.
\item {[58]}
A. K. RAYCHAUDHURI, {\it Theoretical Cosmology} (Clarendon Press, Oxford,
1979).
\item {[59]}
M. DEMIANSKI, $\; \;$R. DE RITIS, G. PLATANIA, P. SCUDELLARO, C. STORNAIOLO,
Phys. Rev. {\bf D 35} (1987) 1181.
\item {[60]}
M. PILATI, Nucl. Phys.
{\bf B 132} (1978) 138.
\item {[61]}
B. CARTER, Gen. Rel. Grav. {\bf 1} (1971) 349.
\item {[62]}
N. M. J. WOODHOUSE,
J. Math. Phys. {\bf 14} (1973) 495.
\item {[63]}
C. J. S. CLARKE,
Commun. Math. Phys. {\bf 41} (1975) 65.
\item {[64]}
C. J. S. CLARKE, Commun. Math. Phys.
{\bf 49} (1976) 17.
\item {[65]}
S. W. HAWKING, A. R. KING, P. J. MC CARTHY,
J. Math. Phys. {\bf 17} (1976) 174.
\item {[66]}
C. J. S. CLARKE, B. G. SCHMIDT,
Gen. Rel. Grav. {\bf 8} (1977) 129.
\item {[67]}
G. F. R. ELLIS, B. G. SCHMIDT, Gen. Rel. Grav.
{\bf 8} (1977) 915.
\item {[68]}
F. TIPLER, Ann. of Phys.
{\bf 108} (1977) 1.
\item {[69]}
C. T. J. DODSON, Intern. J. Theor. Phys.
{\bf 17} (1978) 389.
\item {[70]}
B. CARTER,
in Recent Developments in Gravitation, Carg\`ese
1978, eds. M. L\'evy, S. Deser (Plenum Press, New York, 1979) 41.
\item {[71]}
C. W. LEE,
Gen. Rel. Grav. {\bf 15} (1983) 21.
\item {[72]}
U. D. VYAS, P. S. JOSHI,
Gen. Rel. Grav. {\bf 15} (1983) 553.
\item {[73]}
C. J. S. CLARKE, F. DE FELICE,
Gen. Rel. Grav. {\bf 16} (1984) 139.
\item {[74]}
U. D. VYAS, G. M. AKOLIA, Gen. Rel. Grav.
{\bf 16} (1984) 1045.
\item {[75]}
R. P. A. C. NEWMAN,
Gen. Rel. Grav. {\bf 16} (1984) 1163.
\item {[76]}
R. P. A. C. NEWMAN,
Gen. Rel. Grav. {\bf 16} (1984) 1177.
\item {[77]}
U. D. VYAS, G. M. AKOLIA,
Gen. Rel. Grav. {\bf 18} (1986) 309.
\item {[78]}
R. P. A. C. NEWMAN, C. J. S. CLARKE,
Class. Quantum Grav. {\bf 4} (1987) 53.
\item {[79]}
L. Z. FANG, R. RUFFINI, {\it Quantum Cosmology} (World Scientific, Singapore,
1987).
\item {[80]}
J. J. HALLIWELL, {\it A Bibliography of Papers on Quantum Cosmology}
(1989), Santa Barbara and MIT preprint, NSF-ITP-89-162.
\item {[81]}
Z. PERJ\'ES, {\it Parametric Spinor Approach to Gravity} (1990),
Central Research Institute for Physics, Budapest preprint.
\item {[82]}
L. DABROWSKI, {\it Group Actions on Spinors} (Bibliopolis, Napoli, 1988).
\item {[83]}
P. BUDINICH, A. TRAUTMAN, {\it The Spinorial Chessboard} (Springer-Verlag,
Berlin, 1988).
\item {[84]}
A. LICHNEROWICZ, Coll\`ege de France, Cours 1960-61.
\item {[85]}
C. J. ISHAM, C. N. POPE, N. P. WARNER,
Class. Quantum Grav. {\bf 5} (1988) 1297.
\item {[86]}
E. AGUIRRE-DAB\'AN, M. GUTI\'ERREZ-L\'OPEZ,
Gen. Rel. Grav. {\bf 21} (1989) 45.
\item {[87]}
J. DIECKMANN, Gen. Rel. Grav.
{\bf 20} (1988) 859.
\item {[88]}
I. R\'ACZ, Gen. Rel.
Grav. {\bf 20} (1988) 893.
\item {[89]}
P. S. JOSHI, Gen. Rel. Grav. {\bf 21} (1989) 1227.
\item {[90]}
F. P. ESPOSITO, L. WITTEN, {\it Asymptotic Structure of Space-Time}
(Plenum Press, New York, 1977).
\item {[91]}
F. J. FLAHERTY, $\; \; \;${\it Asymptotic Behaviour of Mass and
Space-Time Geometry}
(Springer-Verlag, Berlin, 1984).
\item {[92]}
S. PERSIDES,
J. Math. Phys. {\bf 20} (1979) 1731.
\item {[93]}
S. PERSIDES,
J. Math. Phys. {\bf 21} (1980) 135.
\item {[94]}
S. PERSIDES,
J. Math. Phys. {\bf 21} (1980) 142.
\item {[95]}
R. P. A. C. NEWMAN,
Commun. Math. Phys. {\bf 123} (1989) 17.
\item {[96]}
A. ASHTEKAR, V. HUSAIN, C. ROVELLI, J. SAMUEL, L. SMOLIN,
Class. Quantum Grav. {\bf 6} (1989) L185.
\item {[97]}
R. P. A. C. NEWMAN, Gen. Rel.
Grav. {\bf 21} (1989) 981.
\item {[98]}
M. KRIELE,
Class. Quantum Grav. {\bf 6} (1989) 1607.
\item {[99]}
G. T. HOROWITZ, A. R. STEIF, Phys. Rev. Lett. {\bf 64} (1990) 260.
\item {[100]}
J. ISENBERG, J. M. NESTER, in General Relativity
and Gravitation, Vol. I, ed. A. Held (Plenum Press, New York, 1980) 23.
\item {[101]}
L. CASTELLANI, P. VAN NIEUWENHUIZEN, M. PILATI,
Phys. Rev. {\bf D 26} (1982) 352.
\item {[102]}
R. DI STEFANO, R. T. RAUCH,
Phys. Rev. {\bf D 26} (1982) 1242.
\item {[103]}
J. PULLIN, Ann. Physik {\bf 46} (1989) 167.
\item {[104]}
S. GOTZES, A. C. HIRSHFELD, Ann. Phys. {\bf 203} (1990) 410.
\item {[105]}
R. GILMORE, {\it Lie Groups, Lie Algebras and Some of Their Applications}
(John Wiley and Sons, New York, 1974).
\bye